%% file: cluster.tex
\PassOptionsToPackage{protrusion, expansion}{microtype} 
\PassOptionsToPackage{dvipsnames}{xcolor} 
\pdfoutput=1
\documentclass[conference, nofonttune]{IEEEtran}\def\UseIEEETemplate{1}  
\IEEEoverridecommandlockouts

\input{grand-preamble.tex}
\usepackage[style=ieee, citestyle=numeric, sortcites=true, backend=bibtex, maxbibnames=99]{biblatex}
\begin{document}
\title{FT K-means: A High-Performance K-means on GPU with Fault Tolerance}

\input{author_ieee}

\thispagestyle{plain}\pagestyle{plain}

\maketitle
\begin{abstract}
K-means is a widely used algorithm in clustering, however, its efficiency is primarily constrained by the computational cost of distance computing. Existing implementations suffer from suboptimal utilization of computational units and lack resilience against soft errors. To address these challenges, we introduce FT K-means, a high-performance GPU-accelerated implementation of K-means with online fault tolerance. We first present a stepwise optimization strategy that achieves competitive performance compared to NVIDIA's cuML library. We further improve FT K-means with a template-based code generation framework that supports different data types and adapts to different input shapes. A novel warp-level tensor-core error correction scheme is proposed to address the failure of existing fault tolerance methods due to memory asynchronization during copy operations. Our experimental evaluations on NVIDIA T4 GPU and A100 GPU demonstrate that FT K-means without fault tolerance outperforms cuML's K-means implementation, showing a performance increase of 10\%-300\% in scenarios involving irregular data shapes. Moreover, the fault tolerance feature of FT K-means introduces only an overhead of 11\%, maintaining robust performance even with tens of errors injected per second.

\end{abstract}






\input{Sections/Section1-Introduction}

\input{Sections/Section2-RelatedWorks.tex}
\input{Sections/Section3-KMeans_without_FT}
\input{Sections/Section4-KMeans_with_FT}

\input{Sections/Section5-Performance_Evaluation.tex}
\input{Sections/Section6-Conclusion.tex}

\newpage
\renewcommand*{\bibfont}{\footnotesize}
\printbibliography[]


\end{document}

%% file: grand-preamble.tex
\usepackage{graphicx}
\usepackage{xcolor}
\definecolor{B}    {HTML}{2b66d3}   
\definecolor{B2}   {HTML}{003399}   
\definecolor{Bv}   {HTML}{0000EB}   
\definecolor{R}    {HTML}{c9171e}
\definecolor{R2}   {HTML}{d7003a}
\definecolor{INK}  {HTML}{595857}
\definecolor{Y}    {HTML}{f1c40f}
\definecolor{G}    {HTML}{009a00}
\definecolor{GRAY} {HTML}{808080}
\definecolor{MAUVE}{HTML}{9400D1}

\newcommand{\cusz}{{\scshape cuSZ}}

\usepackage{array}      
\usepackage{booktabs}
\usepackage{colortbl}
\usepackage{adjustbox}
\usepackage{multirow}
\usepackage{tabu}       

\input{setting/module_text_func}	
\input{setting/module_text_macros}

\input{setting/module_tikz}

\usepackage{anyfontsize}

\usepackage{amsmath}    
\usepackage{amsfonts}
\usepackage{amsthm}
\usepackage{amssymb}    
\usepackage{mathtools}
\usepackage{bm}

\usepackage{caption}
\usepackage{subcaption}

\usepackage[normalem]{ulem}
\usepackage{stackengine}
\usepackage{rotating}	

\usepackage[outercaption]{sidecap}
\sidecaptionvpos{figure}{b}
\sidecaptionvpos{table}{b}
\hypersetup{
	pdftitle={FT K-means: A High-Performance K-means on GPU
with Fault Tolerance},
	pdfauthor={Wu, Ding, et al.},
	pdfsubject={FT K-means, camera-ready},
	pdfkeywords={GPU; Fault Tolerance; Machine Learning; HPC}
}

\usepackage{amsmath}

\DeclareMathOperator*{\argmin}{arg\,min}
\usepackage{listings}
\usepackage{microtype}

\usepackage{enumitem}
\usepackage[switch]{lineno}
\usepackage{algorithm}
\usepackage{algpseudocode}
\usepackage{multicol}
\usepackage{multirow}
\usepackage{caption}
\usepackage{subcaption}
\usepackage{subcaption}
\usepackage{booktabs} 
\usepackage{siunitx} 
\usepackage{etoolbox}
\usepackage{soul}
\usepackage{cancel}
\usepackage[normalem]{ulem}

\usepackage{caption}

\usepackage{amsmath,algorithm,tabularx}
\usepackage{algpseudocode}

\makeatletter
\newcommand{\multiline}[1]{%
  \begin{tabularx}{\dimexpr\linewidth-\ALG@thistlm}[t]{@{}X@{}}
    #1
  \end{tabularx}
}
\renewcommand{\fnum@figure}{Figure \thefigure}
\newcommand{\shixun}[1]{\textcolor{black}{#1}}

\makeatother

\microtypesetup{nopatch={footnote}}

\newcommand{\EDIT}[2]{%
    \bgroup%
    \colorbox{B}{\color{white}#1:}\color{B} #2
    \egroup%
}


%% file: setting/module_text_func.tex
\usepackage{url, hyperref}                  
\hypersetup{
    colorlinks, linkcolor=RoyalBlue, citecolor=RoyalBlue, urlcolor=RoyalBlue,
    allcolors=RoyalBlue,
    pdfborder={0 0 0}}

\usepackage{enumitem}                       
\usepackage{lipsum}                         
\usepackage[plain]{algorithm}       		
\floatstyle{plaintop}
\usepackage{algpseudocode}
\algrenewcommand{\alglinenumber}[1]{{\scriptsize\bfseries\ttfamily\color{R}#1}}
\floatstyle{plaintop}
\restylefloat{algorithm}

\usepackage{xpatch}
\makeatletter
\xpatchcmd{\algorithmic}{\ALG@tlm\z@}{\ALG@tlm\z@\leftmargin 10pt}{}{}
\makeatother

\usepackage{listings}						
\usepackage{setspace}                       

\usepackage{soul}

%% file: setting/module_text_macros.tex
\ifx\UseACMTemplate\undefined
\else

\fi

\ifx\UseIEEETemplate\undefined
\else

\fi

\colorlet{HLCOLOR}{B}
\colorlet{TableAltColor}{gray!20}

%% file: setting/module_tikz.tex
\usepackage{tikz}
\usetikzlibrary{positioning}        
\usetikzlibrary{shapes.multipart}   
\usetikzlibrary{shapes.geometric}   
\usetikzlibrary{arrows}
\usetikzlibrary{arrows.meta}        
\usetikzlibrary{backgrounds}
\usetikzlibrary{patterns}

\usepackage{pgfplots}
\usepackage{pgfplotstable}
\usepgfplotslibrary{groupplots}

%% file: author_ieee.tex
\newcommand{\CoMark}{$^{\star}$}
\newcommand{\UcrMark}{$^\dagger$}
\newcommand{\UHMark}{$^\S$}
\newcommand{\AnlMark}{$^\ddagger$}

\author{%
\normalsize\null
    Shixun Wu\UcrMark\CoMark,
    Yitong Ding\UcrMark\CoMark,\thanks{\CoMark\,Shixun Wu and Yitong Ding contributed equally to this work.}
    Yujia Zhai\UcrMark,
    Jinyang Liu\UHMark,
    Jiajun Huang\UcrMark,
    Zizhe Jian\UcrMark,
    Huangliang Dai\UcrMark,\\
    Sheng Di\AnlMark,
    Bryan M. Wong\UcrMark,
    Zizhong Chen\UcrMark,
    Franck Cappello\AnlMark
    \\
    \IEEEauthorblockA{\UcrMark University of California, Riverside, CA, US}
    \IEEEauthorblockA{\UHMark University of Houston, Houston, TX, US}
    \IEEEauthorblockA{\AnlMark Argonne National Laboratory, Lemont, IL, US}
    \{swu264,yzhai015, jhuan380, zjian106, hdai022\}@ucr.edu,  dingyitongytd@gmail.com, jliu217@central.uh.edu
    \\ sdi1@anl.gov, bryan.wong@ucr.edu, chen@cs.ucr.edu, cappello@mcs.anl.gov
}

%% file: Sections/Section1-Introduction.tex
\section{Introduction}
The K-means, one of the top 10 algorithms in data mining \cite{wu2008top}, is widely used in image classification, vector quantization \cite{gersho2012vector}, knowledge discovery \cite{fayyad1996advances}, and pattern classification \cite{duda1973pattern}. However, K-means is increasingly vulnerable to transient faults caused by high circuit density,  low near-threshold voltage, and low near-threshold voltage \cite{lutz1993analyzing, nicolaidis1999time,laprie1985dependable}. Oliveira et al. \cite{oliveira2017experimental} demonstrated an exascale
system with 190,000 cutting-edge Xeon Phi processors
still suffering from daily transient errors under ECC protection. Recognizing the importance of this issue, the U.S. Department of Energy has named reliability as a major challenge for exascale computing \cite{lucas2014doe}.

Intel Corporation first documented a transient error leading to soft data corruption in 1978, marking a significant recognition of such faults' impact in academia and industry \cite{may1979alpha}. Subsequently, in 2000, Sun Microsystems reported server crashes and outages at major sites like America Online and eBay back to cosmic ray strikes on unprotected caches \cite{baumann2002soft}. Similarly, Virginia Tech, in 2003, had to dismantle and sell its newly assembled Big Mac cluster of 1100 Apple Power Mac G5 computers due to a lack of ECC protection that resulting the system prone to cosmic ray-induced failures \cite{geist2016supercomputing}. Despite advancements in ECC protection, transient faults continue to challenge system reliability. For example, a simulation by Oliveira et al. of an exascale system with 190,000 advanced Xeon Phi processors revealed daily vulnerabilities to transient errors, even with ECC \cite{oliveira2017experimental}. These faults are not just theoretical concerns; real-world impacts have been recorded by Google, which experienced transient faults causing incorrect outputs in its production environment \cite{hochschild2021cores}. In 2018, faced with the ongoing risk of transient faults on its large-scale infrastructure, Meta launched an internal investigation to seek solutions \cite{dixit2021silent}.
Transient faults may cause either fail-stop errors, which lead to system crashes, or fail-continue errors, which produce incorrect outcomes. While fail-stop errors can often be addressed through checkpoint/restart mechanisms \cite{phillips2005scalable, NEURIPS2019_9015, tao2018improving, tensorflow2015-whitepaper} or algorithmic approaches \cite{chen2009optimal, wu2011fault, hakkarinen2012multilevel, hakkarinen2014fail, chen2008scalable, chen2008extending}, fail-continue errors are more problematic as they silently corrupt the state of applications and result in incorrect outputs \cite{mitra2014resilience, cher2014understanding, dongarra2011international, calhoun2017towards, snir2014addressing}. These errors are particularly dangerous in environments where safety is critical \cite{li2017understanding}. This paper focuses on fail-continue errors within the computational logic units, assuming that memory errors and those causing system stops are managed using error-correcting codes and checkpoint/restart techniques. We refer to these issues as soft errors. 

Existing fault tolerance methods are feasible for K-means clustering. Taamneh \cite{taamneh2020parallel} proposes a checkpointing strategy, which involves periodically saving the centroids to stable storage during normal operation and restarting from checkpointed centroids in the event of a failure. However, this method cannot detect silent errors and requires recomputation after an error. Dual modular redundancy (DMR) verifies computational correctness by replicating instructions and comparing the results \cite{oh2002error, oh2002control, reis2005swift, yu2009esoftcheck, chen2016simd}. Figure \ref{fig:kmeans_workflow} left illustrates the K-means workflow. We find that DMR can protect the memory-bound routine efficiently, such as the centroids update phase in Figure \ref{fig:kmeans_workflow} left step 3. It is because the memory latency of loading the data points is so high that all arithmetic operations can be duplicated without introducing an overhead over $1\%$, even for synchronous instructions like atomicAdd. However, DMR failed to protect the compute-bound distance calculation in Figure \ref{fig:kmeans_workflow} left step 1 due to the redundant computations. In Figure \ref{fig:kmeans_workflow} left step 2, nearest cluster matching is fused within the distance computation kernel to eliminate redundant memory access. Algorithm-based fault tolerance (ABFT) reduces this redundancy by using checksums based on equivalence relationships, lowering the redundancy to $O(1/N)$, where $N$ is the problem size \cite{chen2008algorithm}. Efforts have been made to apply ABFT on GPUs, designing kernel fusion schemes to minimize the overhead of ABFT by hiding it within the memory transactions and computing unit gaps \cite{ding2011matrix, zhai2021ft, kosaian2021arithmetic, wu2023anatomy}. 


 However, existing ABFT fusion strategies lack architecture-aware optimization. Kosaian \cite{kosaian2021arithmetic} proposes an error detection scheme for tensor-core but the error correction requires a time-redundant recomputation. Although error correction is discussed in \cite{zhai2021ft,wu2023anatomy}, they only utilize a register-reusing which becomes outdated for Amepere architecture. The register reusing stage is illustrated in the SIMT GEMM part of Figure \ref{fig:kmeans_workflow} and the asynchronous copy is presented in the Tensor-core part. Before Ampere architecture, the global-to-share memory transfer passes through the register file explicitly, enabling register reusing for checksum computation. However, the latest asynchronous memory copy enables a new data path from global to shared memory bypassing the register. Without register reusing, checksum computation requires additional expensive memory read, destroying the carefully designed pipeline. Additionally, these fault tolerance strategies consider matrix multiplication individually, with no awareness of the underlying K-means routine.

Furthermore, in the state-of-the-art K-means implementation, cuML suffers from fixed kernel parameters, resulting in sub-optimal and low utilization of peak computing performance over various problem sizes. For example, the K-means in cuML obtain a performance of less than 10\% of the peak performance for both single and double precisions. Due to the underlying tall-and-skinny matrix multiplication computation inside the K-means, a code generation strategy and a parameter selection scheme are necessary to improve the K-means performance for various of input shapes and data types.

To address these issues, we propose FT K-means, a high-performance K-means implementation equipped with an algorithm-based fault tolerance scheme that detects and corrects silent data corruptions at computing units on the fly. More specifically, our contributions include the following:

 
\begin{figure}[t!]
    \centering
    \includegraphics[width=\linewidth]{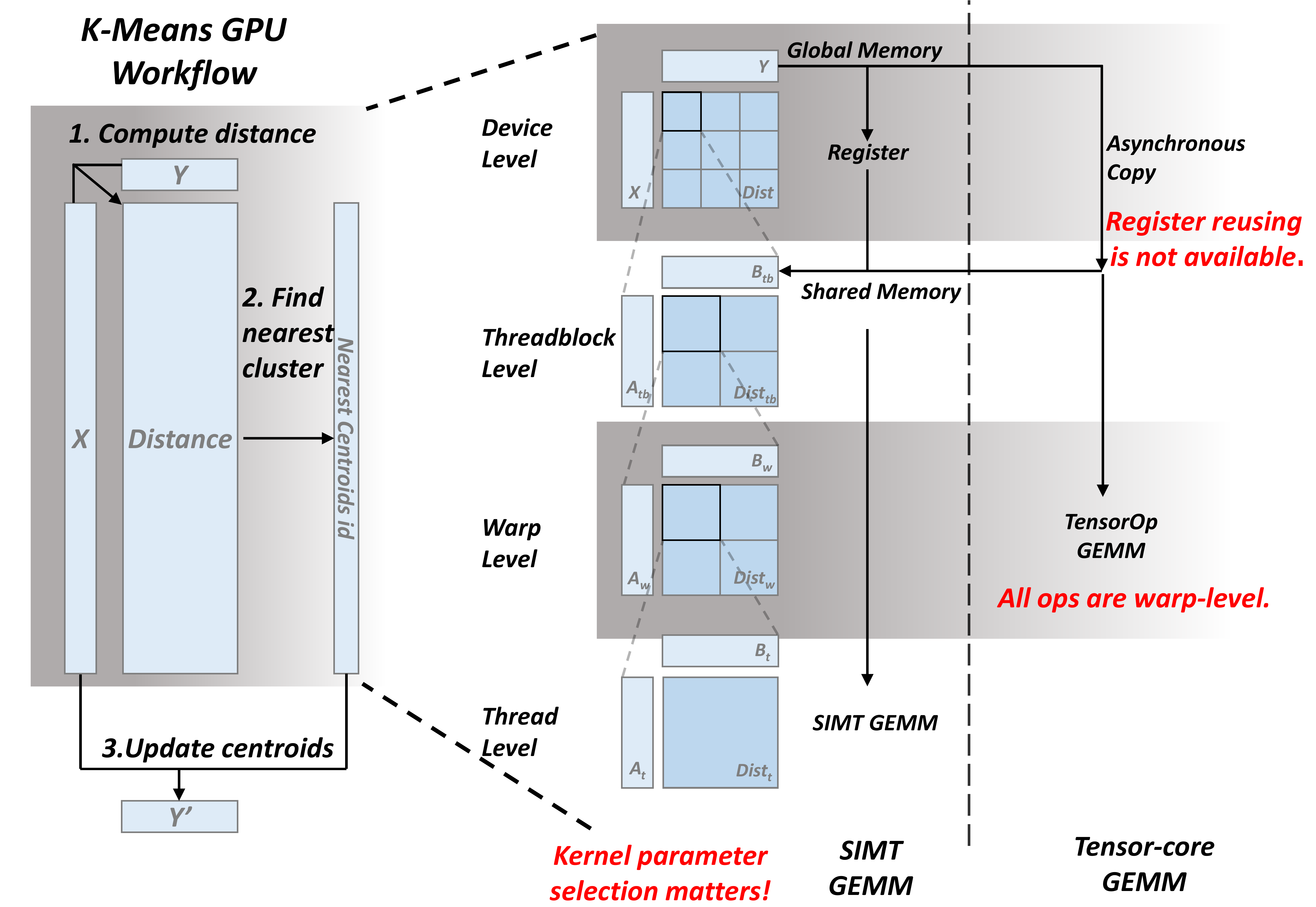}
    \caption{Workflow of K-means. The red illustrates our motivation.}
    \label{fig:kmeans_workflow}
    \vspace{0mm}
\end{figure}

\begin{itemize}[leftmargin=*]

\item We begin our work by optimizing a K-means baseline without fault tolerance. Through a series of optimizations on kernel fusion, FT K-means offers performance competitive to or faster than the state-of-the-art library, cuML. FT K-means is available at an anonymous link.\footnote{https://github.com/shixun404/FT\_KMeans.git} 


\item We explore the fault tolerance K-means designs at the warp level using both the CUDA cores and tensor cores. The combination presents a low overhead even under tens of errors injected per minute. 

\item A template-based code generation strategy is developed to reduce development costs. The template-based can generate K-means kernels with or without fault tolerance for a wide range of input sizes and data types.

\item  Experimental results of single precision and double precision on an NVIDIA A100 server GPU and a Tesla Turing T4 GPU show that FT K-means offers a $10\%-300\%$ improvement compared to the state-of-the-art library cuML. The fault tolerance scheme in FT K-means introduces an average overhead of 11\%, even under tens of error injections per minute. 
\end{itemize}

%% file: Sections/Section2-RelatedWorks.tex
\section{Background and Related Works}\label{sec:background}

 The K-means algorithm aims to categorize \shixun{$M$} objects $\{X_i\}_{i=1}^M$ based on $N$-dimensional features by grouping similar ones into $K$ clusters. Starting with initial centroids ${Y^{(0)}_i}_{i=1}^K$, the algorithm iteratively performs two main steps until satisfies a termination condition. Firstly, it assigns each object $X_i$ to the nearest cluster $Y_j$ using \begin{equation}\argmin_{j=1,\cdots,K}\|X_i - Y_j\|_2.
 \end{equation} Secondly, it updates the centroids of each cluster with  
 \begin{equation}
 Y'_j = \frac{1}{|S_j|} \sum_{i \in S_j} X_i
 \end{equation}
where $ S_j $ represents the set of all points $ X_i$ that are assigned to cluster $ j $, and $|S_j|$ denotes the number of points in cluster $j $. This formula computes the new centroid $Y_j$ as the mean of all points assigned to the cluster $ j $. 
 
\subsection{Fault Model}

FT K-means is designed to identify and correct errors in computing units that could influence the final results. It operates under the assumption that memory errors are managed by ECC \cite{bird2017neutron} and communication reliability issues are addressed by FT-MPI \cite{fagg2000ft}. For compute errors during runtime, a fault-tolerant strategy is implemented based on the single-event upset (SEU) assumption \cite{binder1975satellite,petersen2013single,binder1975satellite}, which posits that only one soft error occurs within each detection and correction interval. This assumption is supported by the low frequency of multiple soft errors in \cite{reis2005swift, zhai2021ft, wu2023ft, ding2011matrix, wu2014ft}. The fault tolerance method involves high reliability in detecting faults with minimal false alarms \cite{turmon2000software}. Specifically, each threadblock randomly selects an element to corrupt by flipping a single bit, either in its 32-bit float representation or 64-bit double representation. A checksum test with a defined threshold $\delta$ then attempts to operate the corrupted computations, followed by an application of an error correction scheme.

\subsection{Previous Work for K-means without Fault Tolerance}

Historically, the simplicity of the k-means algorithm has led to its frequent reimplementation on GPUs. These implementations vary, starting with early GLSL-based versions \cite{shalom2008efficient} and progressing to initial and more recent CUDA versions \cite{farivar2008parallel,li2013speeding}. Lutz et al. \cite{lutz2018efficient} highlighted the significance of single-pass processing in GPU computations, which avoids the redundant loading of data into GPU caches. They reported a significant performance improvement, with speeds over 50 times faster than contemporary CPU implementations and twice as fast as 'double-pass' strategies that involve separate steps. This method is particularly effective for simultaneously processing multiple small k-means instances by using kernel fusion \cite{krulivs2016efficient}. Meanwhile, Cuomo et al. \cite{cuomo2019gpu} conducted a detailed analysis of the performance costs associated with transferring large datasets between CPU and GPU for specialized processing. The kmcuda package provides a GPU-optimized YinYang K-means algorithm on GitHub \cite{ding2015yinyang}, which enhances processing speed on CPUs but faces significant overhead with GPU parallel processing, and is widely used in data analysis tools like R. Nelson and Palmieri \cite{nelson2019don} emphasized the significance of memory management and synchronization, discussing the trade-offs between using global versus shared memory, and different thread synchronization models that involve memory locking. Martin et al. \cite{krulivs2020detailed}  present a detailed analysis of individual computation steps and propose several optimizations that improve the overall performance. Kaiming et al. \cite{ouyang2023kf} introduced an integrated approach to the critical steps of the K-means algorithm, which significantly enhances performance by reducing unnecessary calculations and memory operations, outperforming the Intel DAAL K-means in both sequential and parallel environments.

However, existing works fail to provide enough optimization toward the tensor-core computing units and the asynchronous memory copy presented in the GPU architecture after SM80. cuML \cite{raschka2020machine}, one of the state-of-the-art machine learning libraries on GPU, has provided a K-means implementation using the latest architecture properties. Despite using tensor core computing units and the latest architectural features, the performance of the cuML K-means implementation remains below its hardware's peak potential due to fixed kernel parameters that do not optimize for different input shapes.

\subsection{Algorithm-Based Fault Tolerance}

Soft error protection algorithms aim to identify and rectify errors that can arise in iterative or computationally demanding applications, such as scientific computing \cite{liu2023cusz,huang2023exploring,jian2024cliz,liu2024high}. The development of these algorithms dates back to 1984, with Huang \cite{huang1984algorithm} first specifically designed for matrix-matrix multiplication. The fundamental concept of these algorithms involves encoding matrices \(X\) and \(Y\) into checksums \(X^c\) and \(Y^r\), respectively. This encoding is achieved through the following equations.

\begin{equation}
X\xrightarrow[]{encode}X^c:=e^T X,
\label{eqn:A_encode}
\end{equation}

\begin{equation}
Y\xrightarrow[]{encode}Y^r:=Ye,
\label{eqn:B_encode}
\end{equation}
where $e$ is a column vector, [$1,1,\dots,1$]. The combinations of input matrix and the encoded checksums, $X'= \left[\begin{array}{c}X \\ e^T X\end{array}\right]$ and $Y'=\begin{bmatrix}Y & Y^r\end{bmatrix}$, are then multiplied to get a matrix $D'$ that contains both the correct result and checksum information: 
\begin{equation}
D' = X'Y'= \begin{bmatrix}D & De\\ e^TD & \end{bmatrix} = \begin{bmatrix}D & D^r\\ D^c & \end{bmatrix}.
\label{eqn:C}
\end{equation}
The final accuracy of matrix multiplication can be confirmed by comparing the matrix \(D\) with its checksum counterparts \(D^r\) and \(D^c\). If the difference exceeds a set threshold, it signals an error in the computation. The cost of this checksum encoding and verification is minimal compared to the matrix multiplication itself, offering a cost-effective error detection method. Verification can occur either during (online) or after (offline) the computation.
Chen et al. \cite{chen2008algorithm} introduced an outer-product matrix-matrix multiplication algorithm that maintains the checksum relationship throughout the accumulation process:

\begin{equation}
D' = \sum_iX'(:,i)\cdot Y'(s,:) = \sum_i\begin{bmatrix}D_i & D_ie\\ e^TD_i & \end{bmatrix}.
\end{equation}
where, \(i\) represents the step number in the outer-product update of the matrix \(D\), with \(D_i\) indicating the result of each outer-product, namely \(X'(:, i) \cdot Y'(i,:)\). The offline version of the double-checksum approach can correct only one error in the entire computation. In contrast, the online version corrects a single error at each step of the update, allowing it to address multiple errors throughout the entire program. Ding et al. \cite{ding2011matrix} first presented an implementation of the outer product ABFT-GEMM for GPUs. To minimize the memory latency from checksum operations, Zhai et al. developed combined compute kernels for GEMM on AVX-512-enabled CPUs \cite{zhai2021ft,zhai2023ft}. Kosaian and Rashmi \cite{kosaian2021arithmetic} present a warp-level ABFT implementation capable of error detection, but not correction. Shixun et al. \cite{wu2023anatomy} proposed a fully-fused ABFT-GEMM that effectively detects and corrects computational errors. However, the kernel fusion strategy relies on reusing registers during transfers between global and shared memory. This approach does not extend well to GPU architectures post-Turing, e.g. Ampere, due to the introduction of asynchronous copy instructions. Besides matrix multiplication, the ABFT scheme is widely adopted in fast Fourier transforms \cite{liang2017correcting, wu2024turbofft}, sorting \cite{li2019ft}, iterative methods \cite{chen2013online, tao2016new, chen2014extending}, and computer vision \cite{zhao2020algorithm}.

%% file: Sections/Section3-KMeans_without_FT.tex
\section{K-means without fault tolerance}
\label{sec:design_and_optimizations}
\shixun{The K-means process in each iteration can be divided into two stages. First, for every sample, assign a cluster that has the closest Euclidean distance to it. Second, calculate the geometric center for all samples belonging to a cluster as the new coordinates for this cluster.} \shixun{Consider $M$ samples and $K$ clusters, where each sample has a dimension of $N$.} The time complexity of the first stage is $O(MNK)$, while the time complexity of the second stage is $O(MN)$, so the major bottleneck lies in the first stage. \shixun{In this section, we present the step-wise optimizations of K-means. The optimizations include applying GEMM to K-means, kernel fusion with thread-wise and thread block-wise reduction, broadcast between thread blocks, and enabling tensor core in GEMM.}
\subsection{K-means Stepwise Optimization}
\subsubsection{Basic implementation}
For the cluster assignment stage, we launch a kernel. \shixun{Each thread in this kernel handles a line in the sample matrix (refer to the matrix definition in Fig \ref{fig:kmeans_details}), which represents a sample. The thread loads all centroids in the centroid matrix calculates the Euclidean distance between this sample and every centroid, and chooses the one with the smallest distance as its assigned centroid. For the update centroids stage, M kernels are launched in serial. In kernel $i$, each thread processes one sample. If this sample belongs to kernel $i$, then add all dimensions of this sample to kernel $i$'s corresponding dimensions. Finally, launch a kernel to calculate $\frac{sum\ of\  samples\ belongs\ to\ this\ centroid}{number\ of\ samples\ belongs\ to\ this\ centroid} $, and write the answer back to Centroids matrix as new centroids.}

\subsubsection{GEMM based K-means}
\shixun{Due to the property of Euclidean distance, and we only need to find the closest centroid, we can use GEMM to speed up the cluster assignment process. Ignoring the squared root in distance computation, the distance can be computed in three parts: $\Sigma_k{Samples_{ik}^2} + \Sigma_{k}{Centroids_{jk}^2} - 2 \cdot \Sigma_k{Samples_{ik} \cdot Centroids_{jk}}$ As Figure \ref{fig:kmeans_details} Step 1 depicts, the first two parts of this formula can be computed by squaring elements and summing them up in each row. This can be finished by launching two simple kernels. The third part of the formula has the highest time complexity, and it is exactly in the form of GEMM, so we can launch a GEMM kernel to handle this part and put it together with the first two square terms, and write back to GPU memory. Then we need to launch another kernel to reduce over each row to find the closest centroid for each sample. Also, as illustrated in Figure \ref{fig:kmeans_details} Step 3, for the updating centroids stage, launching N kernels is a great waste of time, because, in kernel j, a large number of threads are idle because the sample which is handled by it doesn't belong to centroid j. So we only launch one kernel to handle N centroids all at once. Each thread still deals with one sample, but it uses atomic add to add the values of this sample in every dimension to its assigned centroid and add one to the counter of this assigned centroid (for average operation). And the last kernel of averaging over each centroid remains unchanged. Our optimization boosts the performance to 25x compared to the basic implementation.}

\subsubsection{Kernel Fusion in thread and thread block level}
\begin{figure}[h]
    \centering
    \includegraphics[scale=0.14]{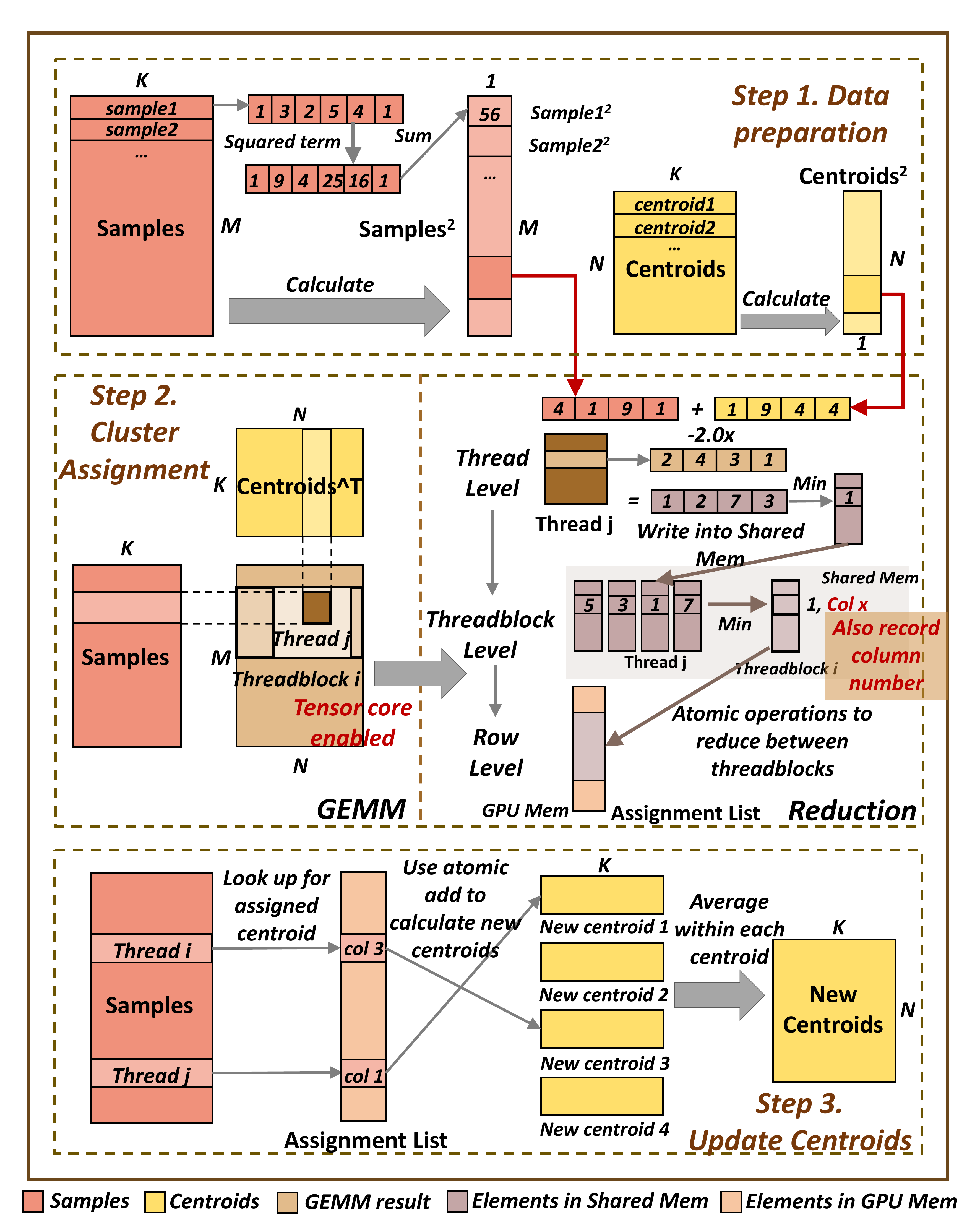}
    \caption{Overview of the optimized K-means.}
    \vspace{0mm}
    \label{fig:kmeans_details}
\end{figure}

After GEMM, we write the result matrix back to GPU memory and load the matrix again in order to do a row-wise reduction. This greatly increases the amount of data movement and increases storage overhead. In this part, we apply kernel fusion \cite{zhai2023bytetransformer,zhai2023architectural} to accomplish part of the reduction within the GEMM kernel. Firstly, as Figure \ref{fig:kmeans_details} Step 2 indicates, each thread in the GEMM kernel handles a small submatrix of the result matrix. After the computation of GEMM, we can simply find the minimum column in each row within this submatrix, and write it to shared memory. \shixun{After all threads in the threadblock finish this step, thread 0 reads all results in shared memory, calculates another row-wise minimum for each row as the final partial answer for this threadblock, and writes it to GPU memory. Assuming each threadblock has size $TM \times TN$, the reduction kernel only needs to load and handle $TN/N$ of data compared to the last part, which obtains a speedup of 1.13x compared to the last part.}

\subsubsection{Threadblock level broadcast}
The optimization above cannot fully eliminate the time and space complexity of launching another reduction kernel after the GEMM kernel. So in this part, we attempt to accomplish cluster assignment within the GEMM kernel. \shixun{Owing to the fact that different threadblocks cannot communicate, we use a broadcast vector and atomic operation to ensure that only one threadblock is changing the assignment answer in one row at a time, i.e. each threadblock needs to acquire for the lock of a row before changing the assignment answer in this row. With this method, the speedup increases to 1.04x compared to the last part.}

\subsubsection{Enabling tensor core in GEMM}
\shixun{Modern Nvidia GPUs are equipped with a tensor core in each Streaming Multiprocessor (SM), designed to accelerate GEMM. Hence we use GEMM kernels in CUTLASS \cite{nv-cutlass} with tensor core and enable TF32 in FP32 precision to further boost performance, instead of our hand-written GEMM kernel.} The reduction part in the last section is placed into the GEMM kernel as an epilogue. \shixun{With this optimization, we achieve a speedup of 1.45x compared to the previous optimization.} And now, the fully optimized GEMM is presented in Figure \ref{fig:kmeans_details}

\subsection{Automatic Code Generation}
cuML has a highly optimized open-source K-means implementation based on CUTLASS. However, in the cluster assignment stage, it has hard-coded parameters in its GEMM kernel, which can trigger low performance in some input sizes. Moreover, the parameters for a CUTLASS GEMM kernel must be hard-coded in order to pass compile-time checking. So the cost of integrating GEMM kernels with customized parameters becomes unacceptable. We propose a code generation strategy to generate kernels with different parameters while minimizing code length to the greatest extent possible.

\textit{Default naming of kernel parameters} 
A group of kernel parameters in cuML and CUTLASS refers to a set of parameters, threadblock level parameters, warp level parameters, and thread level parameters. Each level is composed of three parameters from each dimension. For example, in warp level, the three parameters are labeled Warp.M, Warp.N, and Warp.K, and they refer to the three dimensions of M, N, and K in GEMM.

\begin{figure}[t]
    \centering
    \includegraphics[scale=0.14]{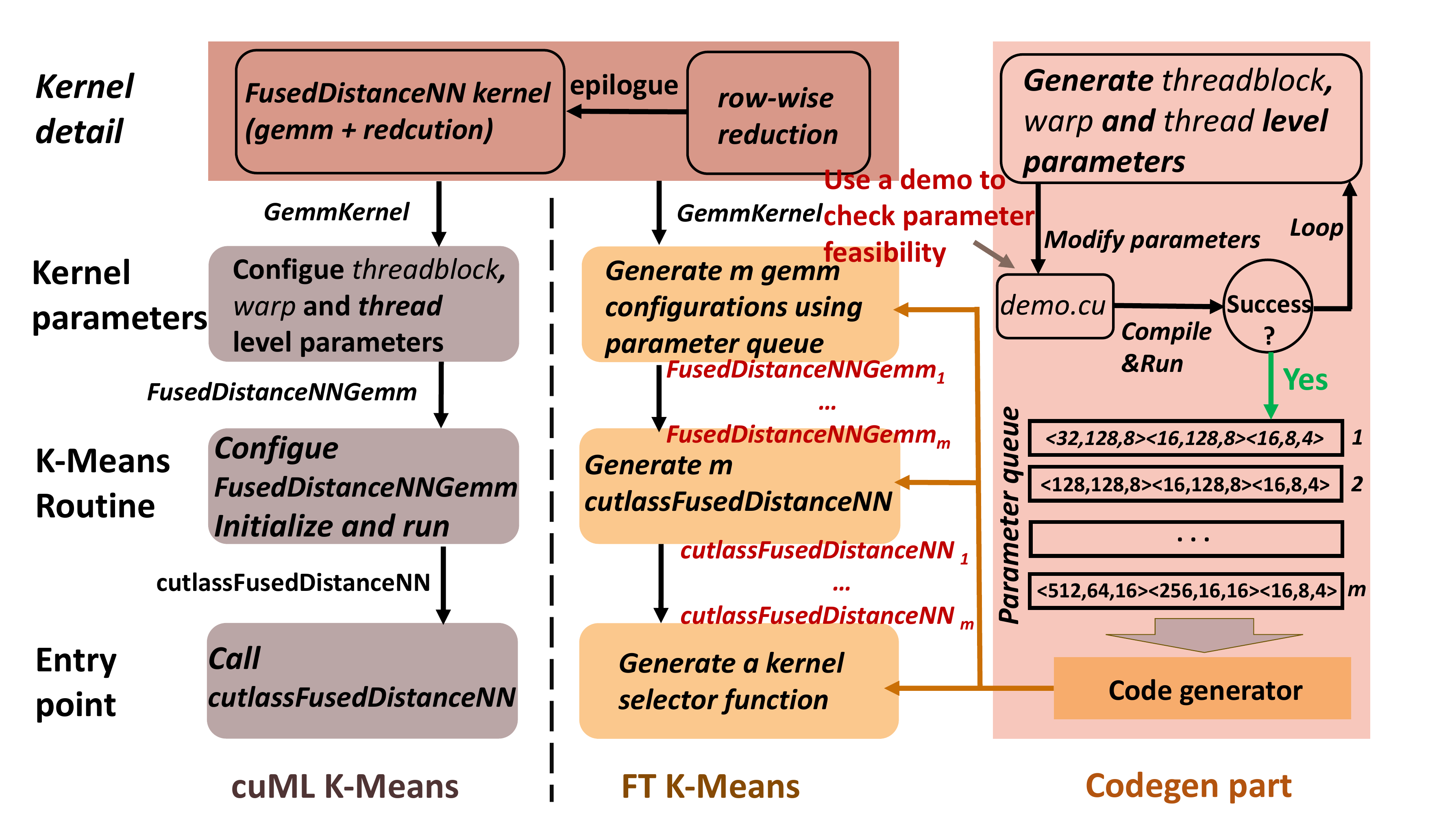}
    \caption{\shixun{Overview of the code generation Strategy}}
    \vspace{0mm}
    \label{fig:codegen}
\end{figure}

\textit{Code Generation Strategy} 
Figure \ref{fig:codegen} demonstrates the method we used in our code generation strategy. The code structure of the CUTLASS GEMM kernel in cuML is on the left side of the figure. Firstly, the epilogue is integrated into the FusedDistanceNN kernel. The FusedDistanceNN kernel handles both GEMM and reduction. And then the kernel is wrapped into GemmKernel. With threadblock, warp, and thread level parameters set, it is then transferred to the next template as FusedDistanceNNGemm. Finally, with all K-means routine sets, the whole cutlassFusedDistanceNN template function works as an interface of the cluster assignment stage. In order to generate a set of feasible kernels with customized parameters, we write a code to test all possible parameters in the search space defined by ourselves, as shown in the code generation part. And for every group of parameters, try it in a demo code. If it can compile and run, which means it is functionally correct. And then we put it in the parameters queue. With all parameters, we then run a code generator to modify the three parts of the original cuML source code, injecting corresponding functions for each parameter in each stage.

\subsubsection{Kernel Parameters}
The kernel parameters used in code generation is not chosen by brute forcing every possible integer in a range for every parameter in the parameter group (a parameter group means: threadblock level, warp level, and thread level parameters). We follow some rules. 1) all parameters must be power of 2 2). Warp.K = Threadblock.K. 3). warp size/thread size is 8 or 16. 4). thread size is fixed for FP32 (16, 8, 4) and FP64 (8, 8, 4) owing to the size of the tensor core.
\subsubsection{\shixun{Code Generation Template}}
\begin{figure}
    \centering
    \includegraphics[width=\linewidth]{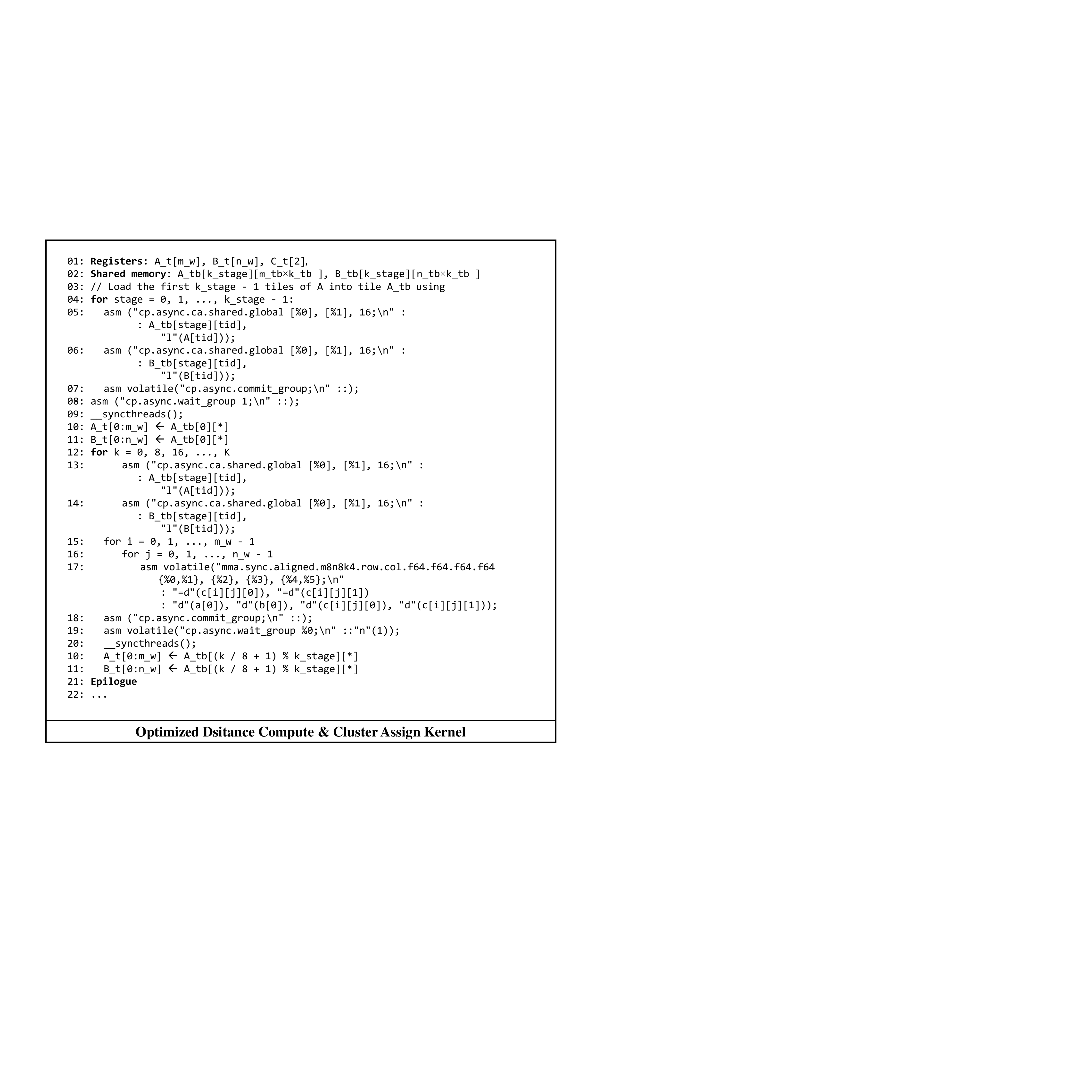}
    \caption{\shixun{Pseudocode: K-means w/o FT}}
    \label{fig:noFT_code}
\end{figure}
\shixun{The code generation strategy defines the K-means kernel using different tiling parameters. For single precision, 157 kernels with different parameter sets are defined while 145 kernels are defined for double precision. The test workflow illustrated in Figure \ref{fig:codegen} checks the feasibility of those kernels and performs the benchmark over 64 problem sizes. The benchmark result of different kernels will be employed as the kernel selection criterion. Below, we give a brief explanation of the K-means kernel so that we can switch to the discussion of the fault tolerance part.}

\shixun{The main body of the kernel is shown in Figure \ref{fig:noFT_code}. In Figure
 \ref{fig:noFT_code}, $A$ stands for samples $X$, $B$ for centroids $Y$, and $C$ for the distance matrix $D$. From line number 04 to 07, an asynchronous multi-stage pipeline from global to shared memory starts and a group barrier is committed for each iteration. At line 08, the kernel waits for at least one group to be ready. Once there the group is prepared, the whole thread block loads data from shared memory into the register synchronously. Next comes the main loop along the number of clusters $K$. At the start of the main loop, the latest memory asynchronous copy is pushed into the pipeline, as shown in lines 13 - 14. After that, the warp-level matrix multiplication is performed by tensor-core units. At the end of this iteration, the latest group is committed and new data is refreshed into the register once a previous group is finished. When the main loop is finished, the memory-bound epilogue performs a reduction along the row of the C matrix. Due to limited space, we skip the detailed description and the reader is welcome to our open-sourced implementation.}

%% file: Sections/Section4-KMeans_with_FT.tex
\section{FT K-means With Fault Tolerance}
\label{sec:fault_tolerant}
In this section, we present the fault tolerance scheme used in FT K-means. Our discussion concentrates on distance computing and the reduction operation to find the nearest cluster. As mentioned before, the centroids updating stage can be handled with a negligible overhead of less than $1\%$ by simply applying the DMR strategy.
\subsection{\shixun{Online correction with location encoding}}
\begin{figure}
    \centering
\includegraphics[width=\linewidth]{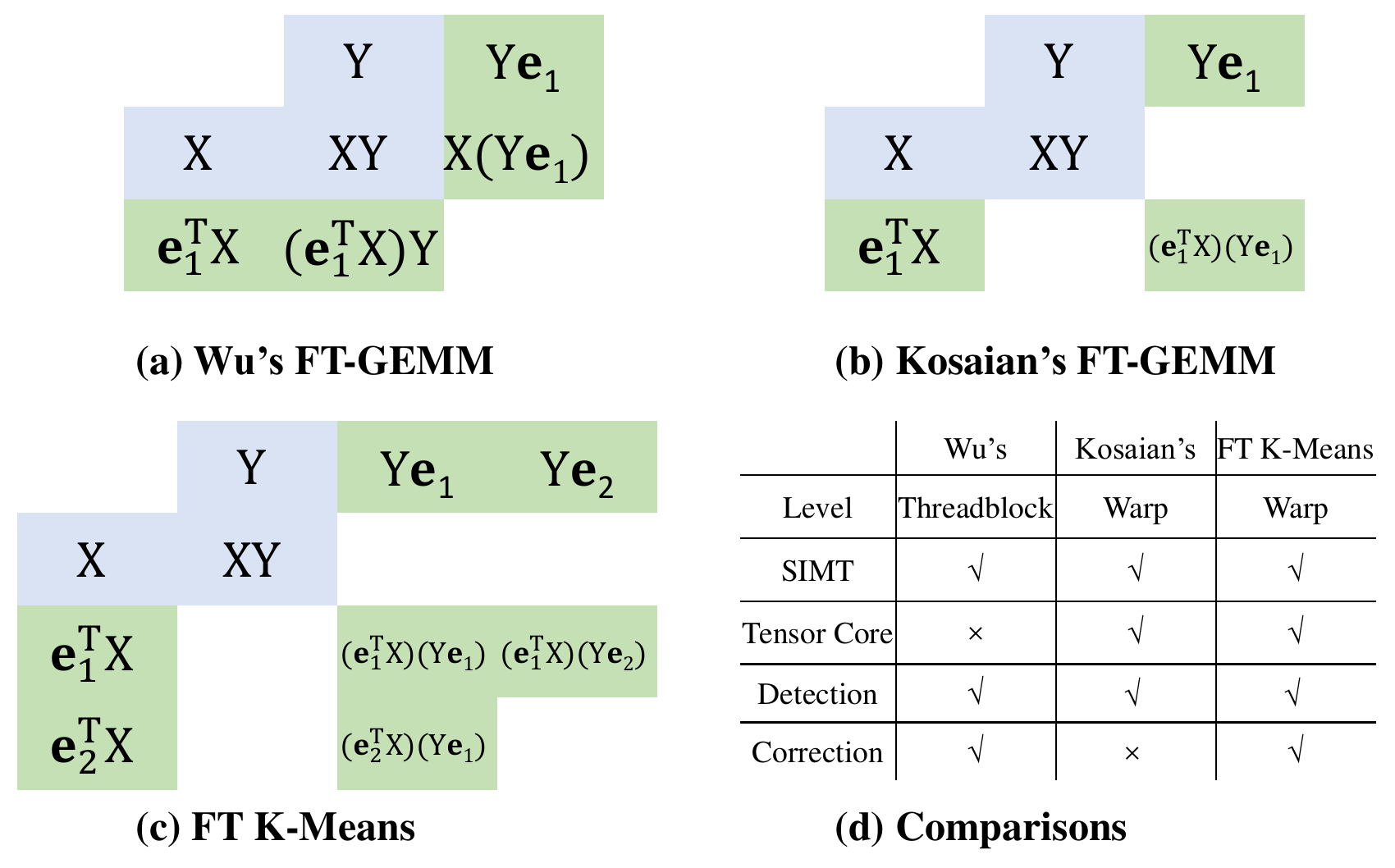}
    \caption{Comparison of SOTA ABFT-GEMM schemes. Blue and green areas are baseline and ABFT computations.}
    \label{fig:comparison_table}
\end{figure}

Figure \ref{fig:comparison_table} compared the fault tolerance scheme in FT K-means with existing state-of-the-art methods in detail. 

One ABFT is illustrated in Figure \ref{fig:comparison_table} (a). To minimize the overhead associated with ABFT, encodings are applied at the thread block level. Aiming to avoid additional latency during GEMM accumulation, the prefetching stage in GEMM is fused with all encodings. In GPU architecture prior to Ampere, the prefetching stage incorporates all ABFT encodings. With a carefully designed prefetching strategy, a warp-level reduction can obtain each element in $e^TX$ and $Ye$ without requiring extra global read operations. Subsequently, the prefetching strategy facilitates the availability of $e^TXY$ and $XYe$ encodings within a thread, eliminating the need for thread block-level communication. Ultimately, the target checksums are accumulated via a threadblock-level reduction.

Compared to the detection scheme specifically designed for tensor-core GPU, as shown in \ref{fig:comparison_table} (b), our method employs a vector $e_2=[1,2, \cdots, N]$ to checksum the inputs again, in addition to the previous $e_1 = [1,1, \cdots, 1]$. Our method doubles the computational cost on CUDA cores ($e^TX, Ye$) and triples the computational cost on tensor cores ($e_1^TXYe_2$, $e_2^TXYe_1$, and $e_1^TXYe_1$). 

\subsection{\shixun{Implementation of FT K-means with fault tolerance}}
\begin{figure}
    \centering
    \includegraphics[width=\linewidth]{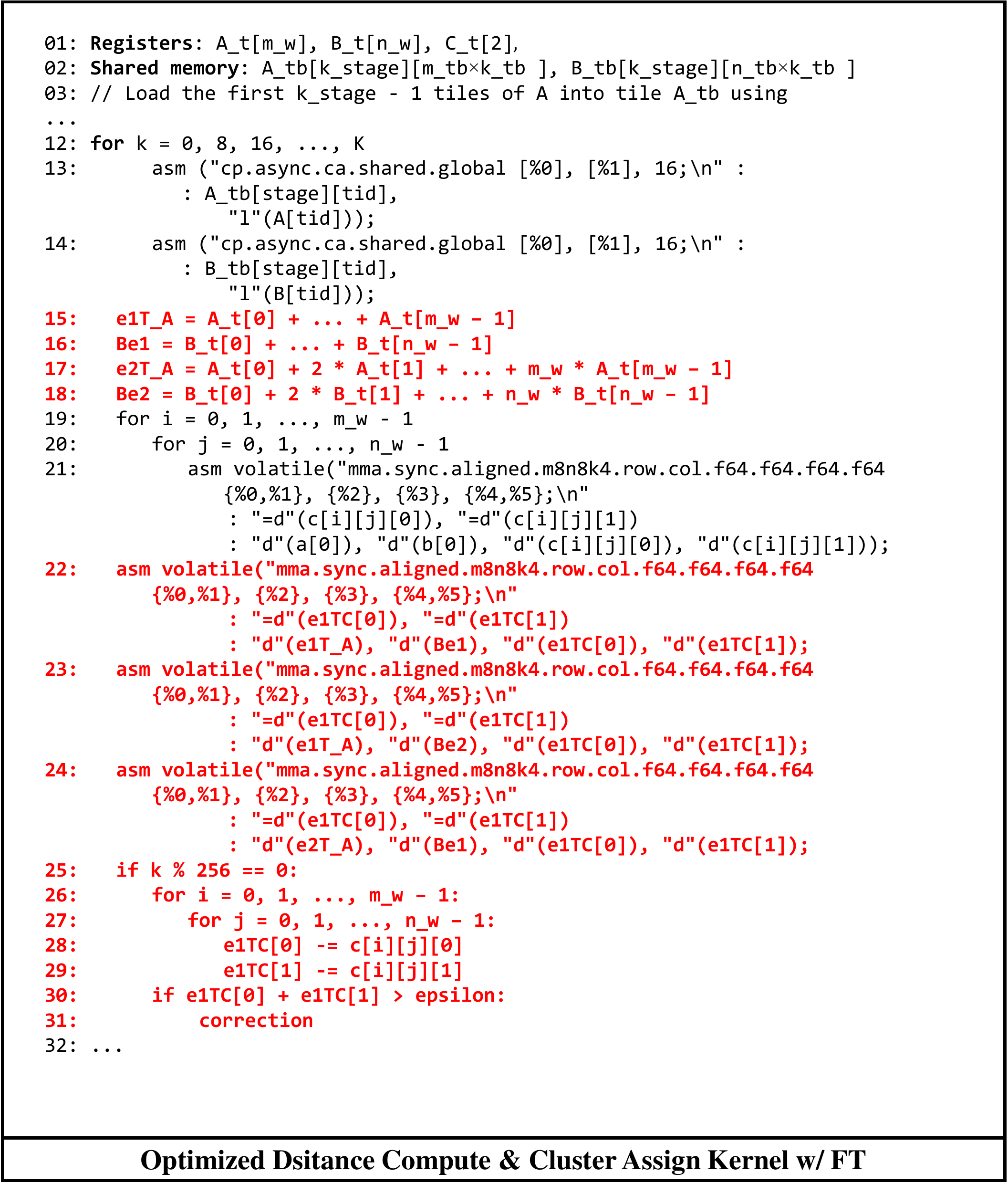}
    \caption{\shixun{Pseudocode: K-means w/ FT}}
    \label{fig:FT_code}
\end{figure}

\shixun{The red part in Figure \ref{fig:FT_code} illustrates the injected instructions to implement FT K-means. From line number 15 to 18, the checksum $e_1^TX, Ye_1, e_2^TX$, and $Ye_2$ are computed. This accumulation takes place inside a thread, getting rid of inter-thread communication and additional memory operation. From line number 22 to 24, $e_1^TXYe_1, e_1^TXYe_2,$ and $e_2^TXYe_1$ are computed via tensor-core MMA operation. The overhead of our ABFT method mainly comes from those three MMAs. Theoretically, they will incur a computation overhead of $\frac{3}{m_w\times n_w}$. Assume $m_w = 4$ and $n_w=2$, the theoretical overhead is $37.5\%$. However, from our experimental evaluation in Section \ref{sec:results}, the overhead is only $11\%$ on average. This theory-experiment mismatch indicates that at the thread level and warp level, there remains a 27.5\% execution bubble between computation and memory, which is available for kernel fusion. Actually, we first try to get the checksums $e_1^TX, Ye_1, e_2^TX$, and $Ye_2$ using the tensor core as well, namely embedding $e_1, e_2$ into a new matrix operand so that we can get the checksum through several tensor core operation. However, those tensor operations cannot be hidden behind the memory footprint, resulting in a 50\% overhead approximately. }

%% file: Sections/Section5-Performance_Evaluation.tex
\section{Performance Evaluation}
\label{sec:results}

We evaluate our K-means on two NVIDIA GPUs, a Tesla Turing T4 and a 40GB A100-PCIE GPU. The Tesla T4 GPU is connected to a node with two 16-core Intel Xeon Silver 4216 CPUs, whose boost frequency is up to 3.2 GHz. The associated CPU main memory system has a capacity of 512 GB at 2400 MHz. The A100 GPU is connected to a node with one 64-core AMD EPYC 7763 CPU with a boost frequency of 3.5 GHz. We compile programs using CUDA $\mathtt{11.6}$ with the $\mathtt{-O3}$ optimization flag on the Tesla T4 machine, and using CUDA $\mathtt{12.0}$ on the A100 machine. A100 has a peak computational performance of $19.5$ TFLOPS for single precision and $9.7$ TFLOPS for double precision. The memory bandwidth is $1.55$ TB/s. T4 has a peak performance of $8.1$ TFLOPS for single precision and a peak performance of $0.253$ TFLOPS for double precision. The bandwidth of T4 is $320$ GB/s. We first demonstrate the benchmark result between FT K-means without fault tolerance and cuML for FP32 and FP64 on A100. Next, we evaluate the FT K-means under fault tolerance. Then, we benchmark FT K-means under error injections with cuML and Wu's ABFT. Finally, we present the same performance evaluation on the T4 GPU. All experimental results are averaged over ten trials.

\subsection{Benchmarking K-means without Fault Tolerance}
\shixun{In this subsection, we evaluate the performance of FT K-means w/o checksum.}
\begin{figure}[tp]
    \centering
    \includegraphics[width=\linewidth]{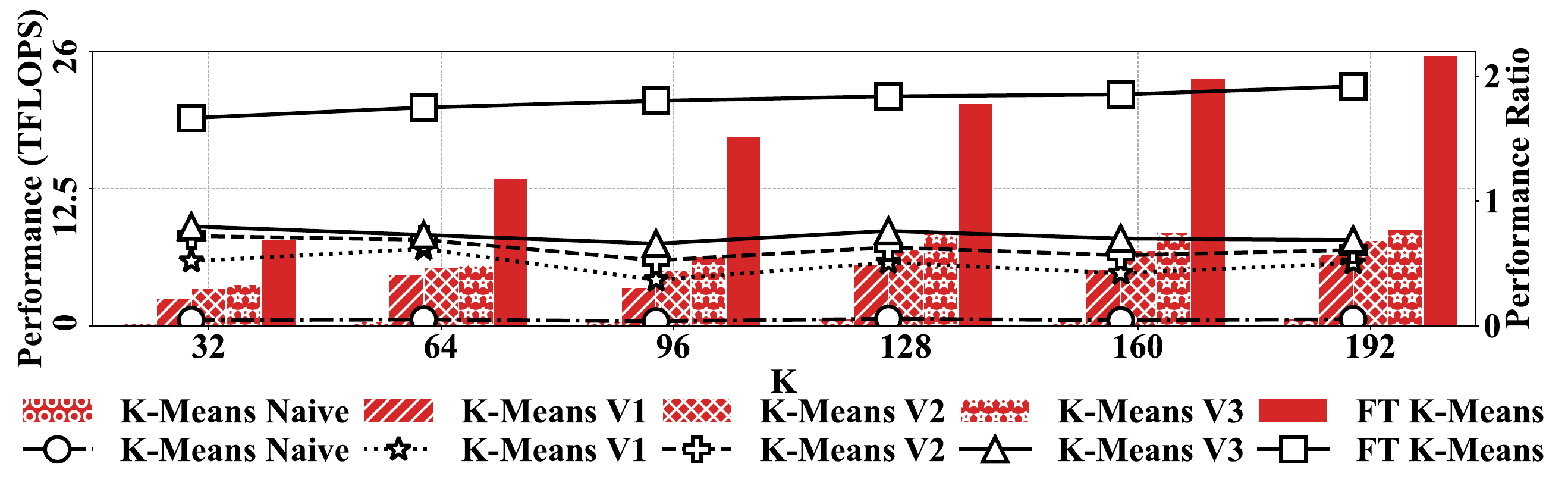}
    \caption{K-means w/o FT stepwise optimizations on A100, FP32 (N=131072, N=128)}
    \label{fig:stepwise_FP32_A100}
    \vspace{0mm}
\end{figure}
\subsubsection{Step-wise optimizations for K-means} 

Figure \ref{fig:stepwise_FP32_A100} demonstrates how our K-means distance kernel is optimized from 5\% to 182\% of cuML stepwise. The performance is measured with GFLOPS (bar plot, the left y-axis) and the performance ratio with respect to cuML (line chart, the right y-axis). Without using any optimizations, the K-means Naive obtains a performance of 482 GFLOPS. Next K-means V1 employs GEMM to K-means distance calculation. The performance improved from 482 GFLOPS to 4662 GFLOPS. Then, K-means V2 adds kernel fusion to both thread and threadblock levels, reducing memory operations. The performance is improved to 5902 GFLOPS. After that, we apply threadblock level broadcast to further reduce memory bound. The performance of K-means V3 achieves 6916 GFLOPS. With tensor core enabled and parameter selection, we finally achieved 17686 GFLOPS, which exceeded the performance of cuML (9676 GFLOPS).

\begin{figure}[ht]
    \vspace{0mm}
    \centering
    \includegraphics[width=\linewidth]{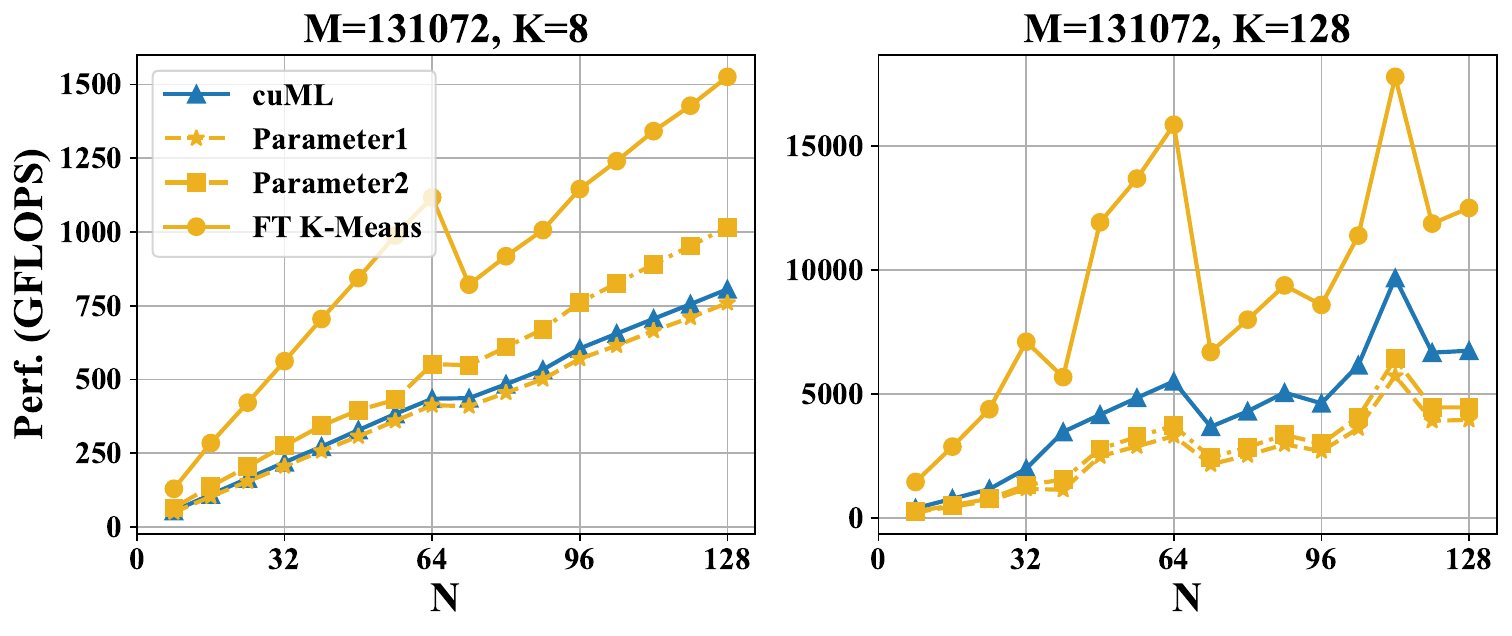}
    \caption{FP32 precision comparison of K-means performance at distance step without fault tolerance with FT K-means, Selected parameters and cuML on an A100 GPU, with M and K fixed.}
    \label{fig:a100_fixMK_fp32}
    \vspace{0mm}
\end{figure}

\begin{figure}[ht]
    \vspace{0mm}
    \centering
    \includegraphics[width=\linewidth]{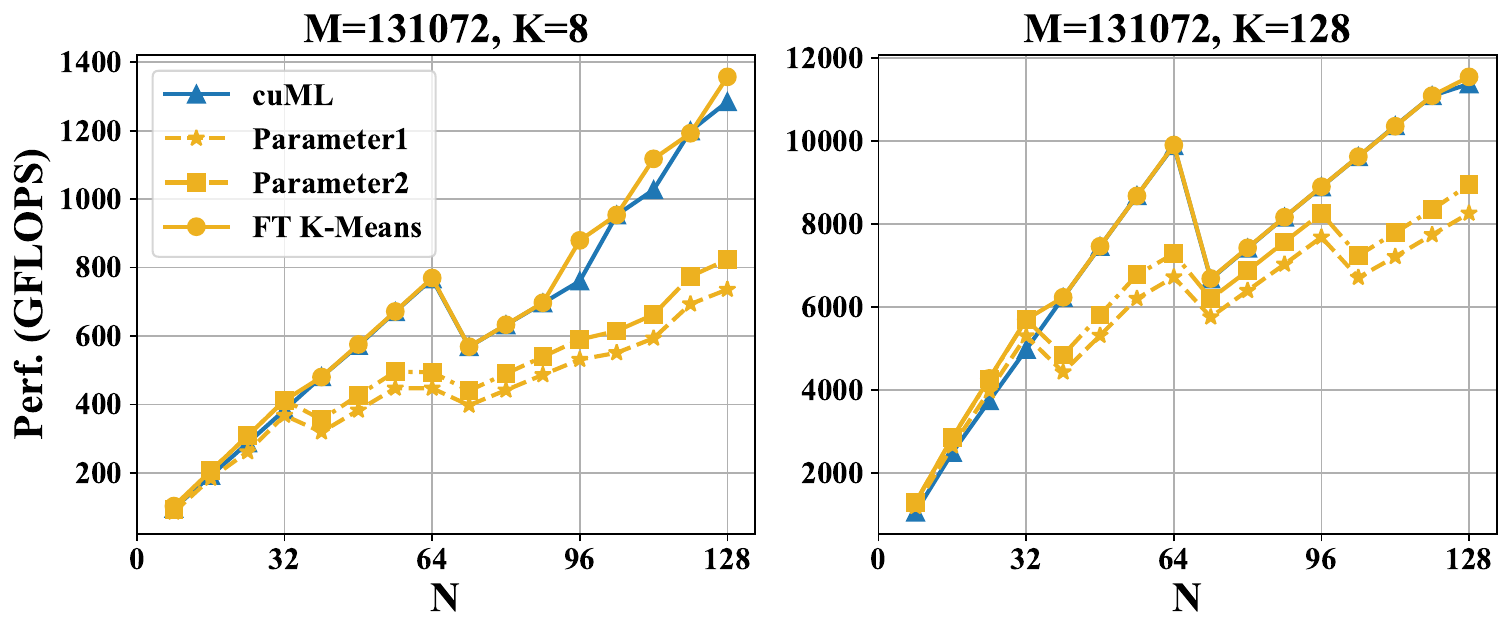}
    \caption{Double precision comparison of K-means performance at distance step without fault tolerance with FT K-means, Selected parameters and cuML on an A100 GPU, with M and K fixed.}
    \label{fig:a100_fixMK_fp64}
    \vspace{0mm}
\end{figure}

\subsubsection{Performance evaluation with M and K fixed} 

Figure \ref{fig:a100_fixMK_fp32} and \ref{fig:a100_fixMK_fp64} demonstrate a performance evaluation in distance step between FT K-means, two selected parameters relative to cuML (all without fault tolerance) in FP32 and FP64 precision when M and K are fixed. We tested the performance of different methods under two values of K, $K=8$ and $K=128$. These are cases representing two situations: one with very few dimensions of data and the other one with relatively more dimensions of data. The selected parameters are labeled Parameter1 and Parameter2, and they are chosen based on experience. The same parameter name refers to different parameters in FP32 and FP64, but their performance is similar. For both FP32 and FP64 precision, parameters selected through experience cannot achieve good performance. Parameter 1 is always slower than cuML, with an average overhead of 15\%. For parameter 2, it slightly exceeded the performance of cuML when $K=8$ in FP32, and it achieves the performance of cuML in some data points where N is small. However, its overall performance is still 5\% slower than cuML. Using code generation strategy, we achieved 235\% speedup compared to cuML under FP32 precision, and the gain in performance is significant even when K is relatively larger. However, under FP64 precision, improvements in performance are minimal, with an overall speedup of 4\% in these two cases. The curves of cuML and FT K-means are almost coincident in a large portion of data points.

\begin{figure}[ht]
\vspace{0mm}
    \centering
    \includegraphics[width=1\linewidth]{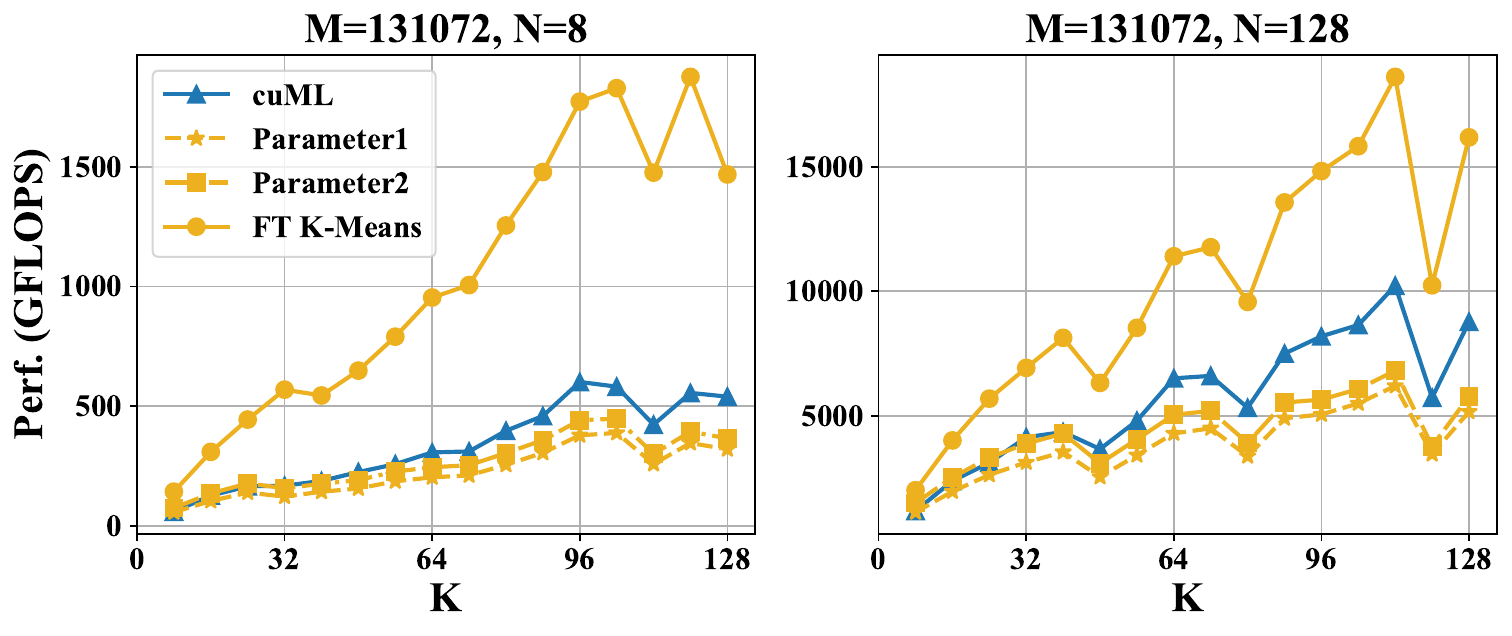}
    \caption{FP32 precision comparison of K-means performance at distance step without fault tolerance with FT K-means, selected parameters, and cuML on an A100 GPU, with M and N fixed}
    \label{fig:a100_fixMN_fp32}
    \vspace{0mm}
\end{figure}

\begin{figure}[ht]
\vspace{0mm}
    \centering
    \includegraphics[width=1\linewidth]{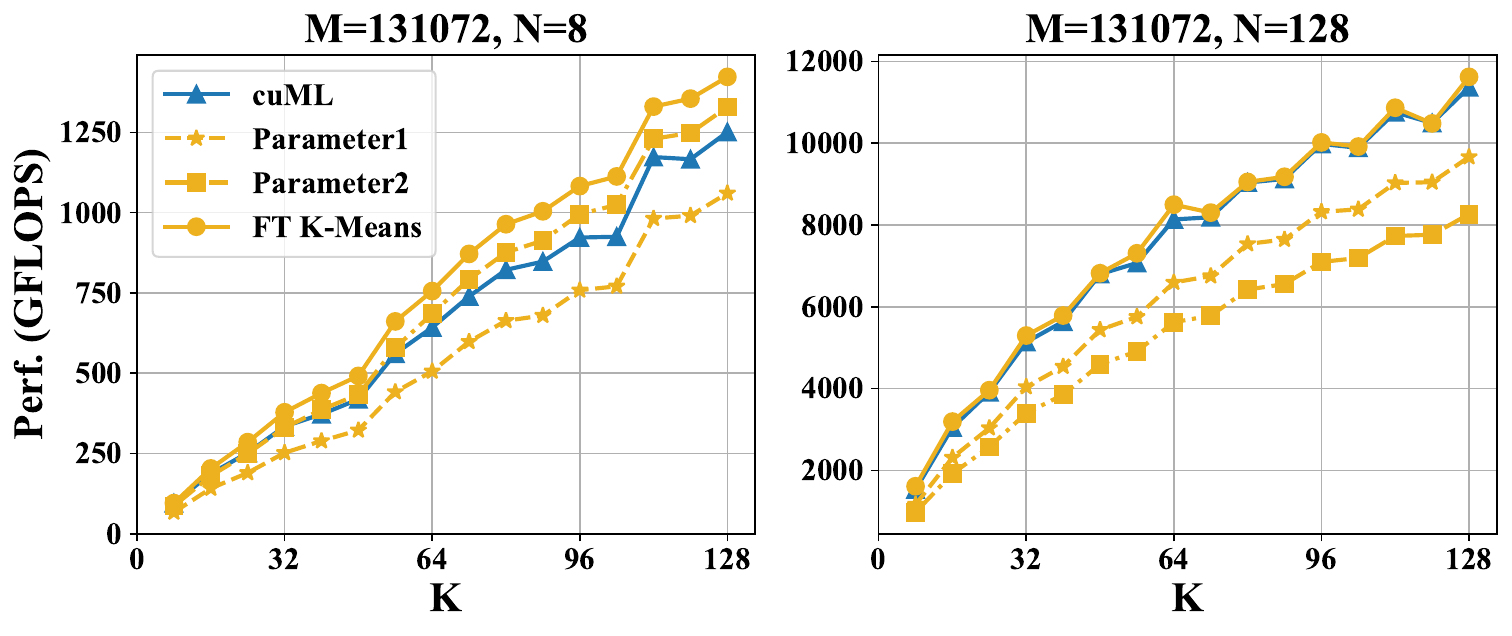}
    \caption{FP64 precision comparison of K-means performance at distance step without fault tolerance with FT K-means, Selected parameters, and cuML on an A100 GPU, with M and N fixed}
    \label{fig:a100_fixMN_fp64}
    \vspace{0mm}
\end{figure}

\subsubsection{Performance evaluation with M and N fixed} 

Figure \ref{fig:a100_fixMN_fp32} and \ref{fig:a100_fixMN_fp64} offer a performance evaluation in distance step between FT K-means, two selected parameters relative to cuML (all without fault tolerance) in FP32 and FP64 precision when M and N are fixed. The main parameter settings are similar to the previous section, and $N=8$ and $N=128$ indicate fewer clustering centroids and relatively more clustering centroids. For both FP32 and FP64 precision, parameters selected through experience cannot achieve good performance. Parameter 1 has an average overhead of 30\%. Parameter 2 exceeded the performance of cuML when K is small in some FP32 cases, and outperforms cuML in FP64 when $N=8$ (1.03x speedup). but its overall performance is still 15\% slower than cuML. Using code generation strategy, we obtained 239\% speedup compared to cuML under FP32 precision, which is similar to fixing M and K. Under FP64 precision, performance improvements are higher than in the previous section, which is 8\% in these two cases. And the speedup in small N is relatively considerable (15\%).

\begin{figure}[ht]
\vspace{0mm}
    \centering
    \includegraphics[width=1\linewidth]{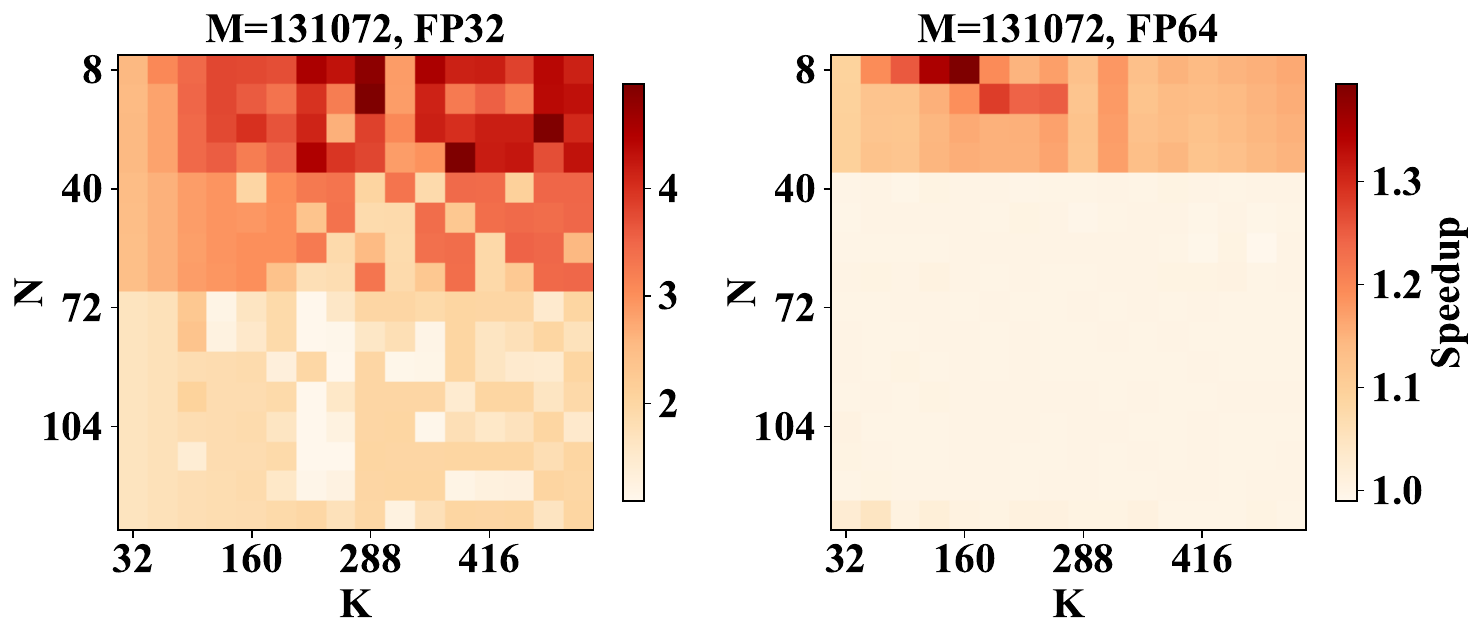}
    \caption{Speedup of FT K-means compared to cuML on FP32 and FP64}
    \label{fig:a100_fixM_heatmap}
    \vspace{0mm}
\end{figure}

\subsubsection{Overall performance evaluation} 
Figure \ref{fig:a100_fixM_heatmap} illustrates an overall comparison between our code generation method and cuML K-means on both FP32 and FP64. For FP32. Our approach shows significant performance improvement, with a maximum speedup of 4.55x and an average speedup of 2.49x. From the perspective of dimension N, the figure indicates a clear trend that the performance improvement diminishes. Furthermore, $N=64$ serves as a clear threshold, beyond which speedup essentially decreases to below 2.0x. The marginal improvement becomes even more common in the case of FP64. The average speedup is only 1.04x, with a maximum speedup of 1.39x. When N exceeds 32, the performance of our method drops to almost identical to cuML. Meanwhile, there is no apparent trend observed in dimension K for both FP32 and FP64.

\begin{figure}[ht]
\vspace{0mm}
    \centering
    \includegraphics[width=1\linewidth]{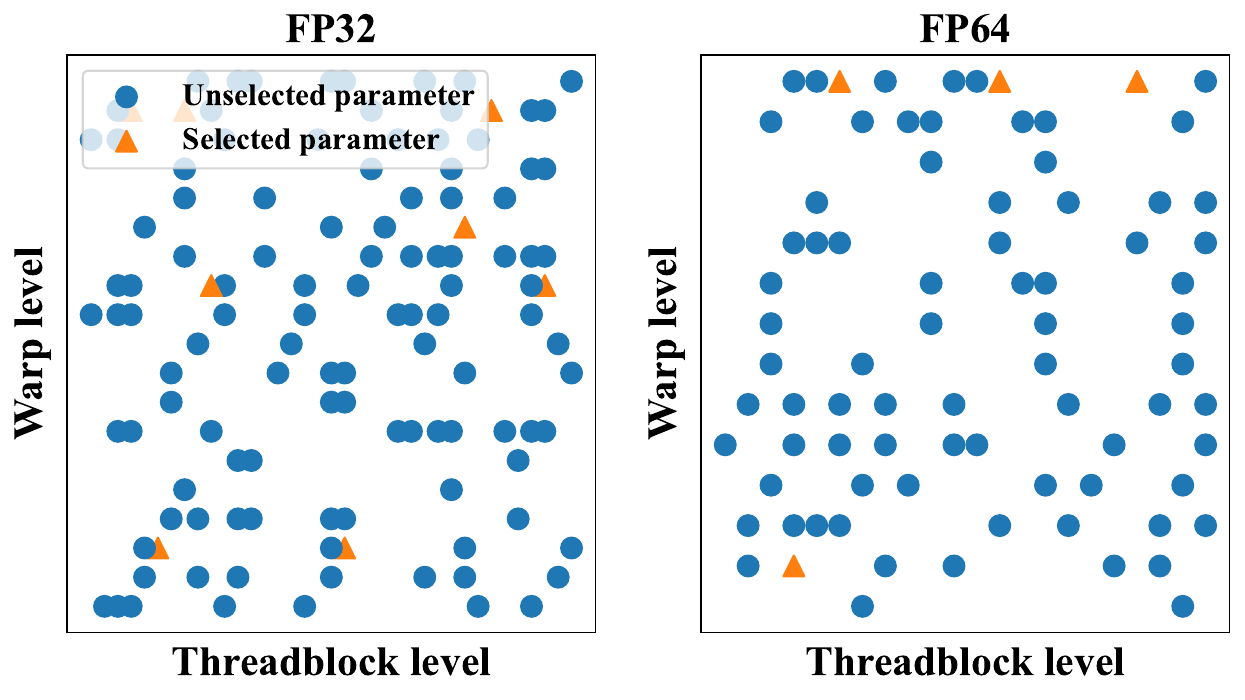}
    \caption{Parameter selection in threadblock and warp level for FP32 and FFP64}
    \label{fig:a100_parameter_selection}
    \vspace{0mm}
\end{figure}

\subsubsection{Evaluation of parameter selection} 
\begin{figure}[ht]
\vspace{0mm}
    \centering
    \includegraphics[width=1\linewidth]{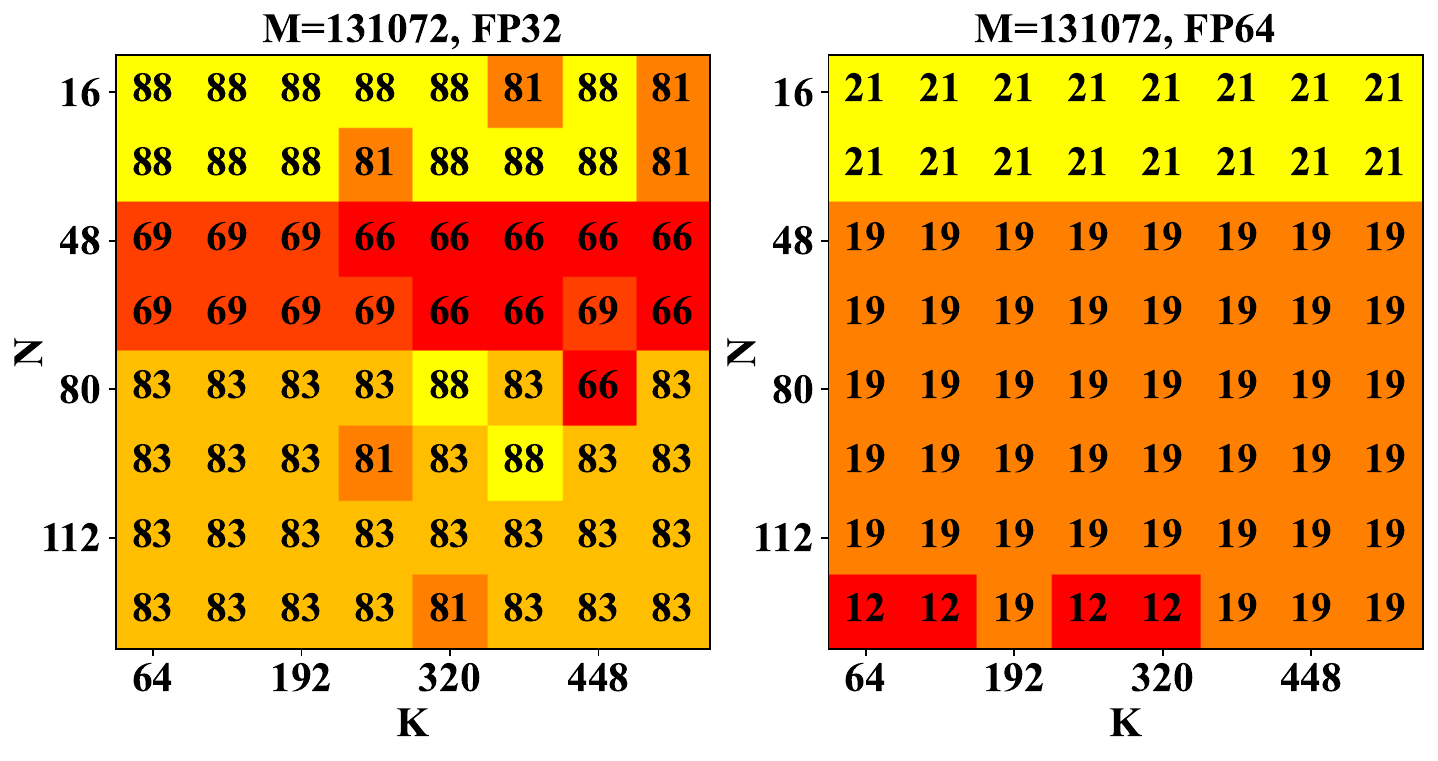}
    \caption{The selected parameter number in FP32 and FP64}
    \label{fig:a100_parameter_occupancy}
    \vspace{0mm}
\end{figure}

The experimental results from previous sections motivate us to analyze the parameters chosen for our code generation method for each data point. For the data size of Figure \ref{fig:a100_fixM_heatmap}, we generated 120 groups of FP32 parameters and 80 groups of FP64 parameters to be selected from. However, only 7 groups of FP32 parameters and 4 groups of FP64 parameters are actually chosen in at least one data point. All threadblock and warp level parameters are shown in Figure \ref{fig:a100_parameter_selection}. And thread level parameters are fixed (FP32: 16,8,4 FP64: 8,8,4) owing to the limited size of the tensor core. Moreover, figure \ref{fig:a100_parameter_occupancy} illustrates the relationship between each data point and its corresponding selected parameters. We will analyze in detail the relationship between the optimal parameters generated by the code generator and the cuML parameters in the next section.

\subsubsection{Detailed analysis of parameters}
For FP32, as shown in figure \ref{fig:a100_parameter_occupancy}, the data points can be divided into three parts: $N \leq 32$, $32 < N \leq 64$ and $64 < N$. When $N \leq 32$, one of the optimal parameters is parameter 88. As table \ref{table:a100_parameters} indicates, compared to cuML, it has a bigger threadblock and warp size in the M direction and a smaller size in the N direction. That is reasonable due to the small N size in these data points. threadblock. N = 256 is too big that more than 87.5\% of the threadblock size contains blank data, so the occupancy is very low. Furthermore, warp.N of cuML, which is 64, exceeds the size of N. Therefore the improvement of our new parameters is significant. As N increases, i.e. $32 < N \leq 64$, the warp size of cuML becomes justified, and our parameter (parameter 69) sticks to this size. However, the threadblock size remains too big in N direction, so a more balanced version of threadblock $64, 128, 16$ outperforms. When N keeps increasing, although the occupancy of cuML achieves a reasonable scope, a more balanced threadblock size, e.g. parameter 83 can reduce the overall amount of data movement from GPU memory to shared memory, and therefore improve performance.

For FP64, the data points only have two main parts: $N \leq 32$ and $32 < N$. When N is relatively small, parameter 21 has an advantage in its small Threadblock. N, and increases occupancy. As N increases, the parameters that are balanced in the M and N direction have better performance. As shown in table \ref{table:a100_parameters}, our parameter 19 is actually identical to cuML's parameter, which demonstrates that the best parameter choice is cuML's parameter in these cases. This illustrates an interesting phenomenon: the parameter choices in FP32 are much more than those in FP64. Moreover, there is also greater potential for performance improvement in FP32 than in FP64. There are several reasons for this.

\begin{table}[ht]
    \centering
    \fontsize{10}{12}\selectfont
    \begin{adjustbox}{max width=\textwidth}
        \begin{tabular}{|c|c|c|c|}
            \hline
            \multicolumn{4}{|c|}{\textbf{FP32}} \\
            \hline
            \multicolumn{1}{|c|}{ID} & \multicolumn{1}{c|}{Threadblock} & \multicolumn{1}{c|}{Warp} & \multicolumn{1}{c|}{Thread} \\
            \hline
            88 & 256, 32, 16 & 64, 32, 16 & 16, 8, 4 \\
            69 & 128, 64, 16 & 32, 64, 16 & 16, 8, 4 \\
            83 & 64, 128, 16 & 64, 32, 16 & 16, 8, 4 \\
            cuML & 32, 256, 16 & 32, 64, 16 & 16, 8, 4 \\
            \hline
            \hline
            \multicolumn{4}{|c|}{\textbf{FP64}} \\
            \hline
            \multicolumn{1}{|c|}{ID} & \multicolumn{1}{c|}{Threadblock} & \multicolumn{1}{c|}{Warp} & \multicolumn{1}{c|}{Thread} \\
            \hline
            21 & 128, 32, 16 & 32, 32, 16 & 8, 8, 4 \\
            19 & 64, 64, 16 & 32, 32, 16 & 8, 8, 4 \\
            cuML & 64, 64, 16 & 32, 32, 16 & 8, 8, 4 \\
            \hline
        \end{tabular}
    \end{adjustbox}
    \vspace{2mm}
    \caption{Partial parameters of FT K-means for FP32 and FP64}
    \label{table:a100_parameters}
\end{table}

\shixun{First, CUTLASS enables TF32 in the FP32 GEMM kernel, which improves the processing speed of tensor cores.} Therefore, the overhead of data movement and epilogue (row-wise reduction) becomes more critical, resulting in greater potential for alternative parameters. Second, the memory alignment requirement for FP64 is more strict than FP32 and is fixed to 1 in CUTLASS's implementation. So the degree of vectorization for FP64 is lower. So a balanced data fetching pattern (Threadblock.M = Threadblock.N) is crucial for increasing performance. Third, the thread-level parameters of FP64 are smaller than those of FP32. Therefore even when N is relatively small, it is easier for a general parameter to obtain high occupancy. In the future, deep reinforcement learning can help accelerate the hyperparameter tuning \cite{johnston2023curriculum,liu2023stationary,johnston2023downlink}.

\subsection{FT K-means with Fault Tolerance}

\begin{figure}[ht]
    \vspace{0mm}
    \centering
    \includegraphics[width=\linewidth]{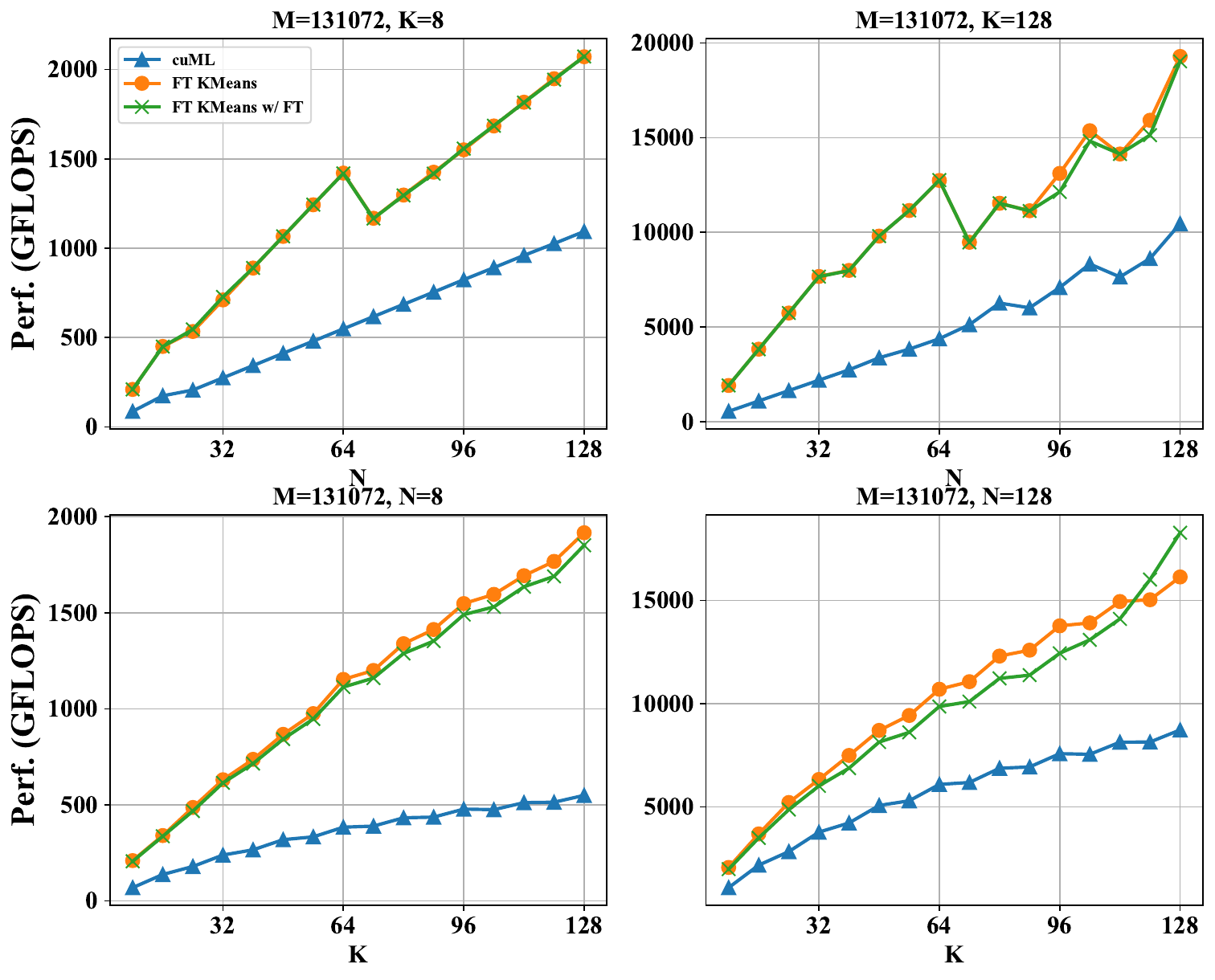}
    \caption{FT K-means with fault tolerance on A100: FP32.}
    \label{fig:a100_fixMK_fp32_ft}
    \vspace{0mm}
\end{figure}

As shown in Figure \ref{fig:a100_fixMK_fp32_ft}, the experimental evaluation of the FT K-means algorithm with fault tolerance demonstrates a remarkably low overhead across various configurations. Specifically, in configurations with \(K=8\) clusters, the overhead was maintained at a minimal \(-0.24\%\), illustrating the negligible impact of integrating fault tolerance mechanisms. Even when the number of clusters was significantly increased to \(K=128\), the overhead remained consistently low at \(1.93\%\). For fixed $N$ scenarios, the overhead was even lower at \(0.96\%\). These results highlight the efficiency of the FT K-means algorithm's fault tolerance, which effectively minimizes additional computational burdens while maintaining robustness across diverse input shapes.

\begin{figure}[ht]
    \vspace{0mm}
    \centering
    \includegraphics[width=\linewidth]{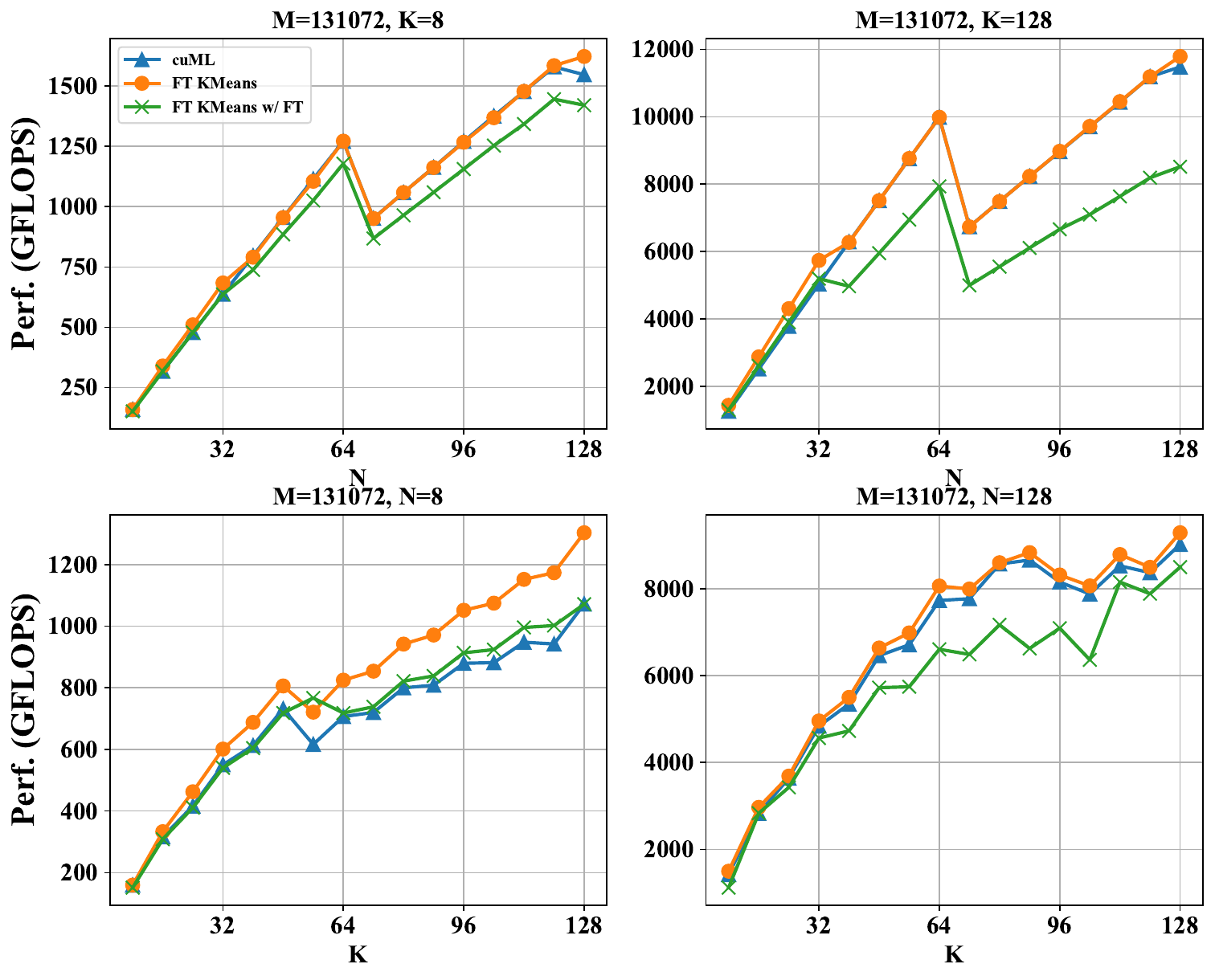}
    \caption{Benchmark FT K-means with fault tolerance: FP64.}
    \label{fig:a100_fixMK_fp64_ft}
    \vspace{0mm}
\end{figure}

In Figure \ref{fig:a100_fixMK_fp64_ft}, FT K-means with fault tolerance 
 presents an average overhead of 13\% for double precision. For small number of clusters (\(K=8\)), the overhead is \(7.9\%\), ranging from \(4.6\%\) to \(12.45\%\). It suggests that the algorithm maintains consistent performance even when computational precision requirements are increased. When the number of clusters increased to \(K=128\), the overhead results in \(20\%\), indicating the performance bottleneck changes to computing. For input shapes of fixed N, \(N=8\) and \(N=128\), the overhead was reduced to \(0.8897\%\). These results underscore the FT K-means algorithm's ability to efficiently manage fault tolerance with minimal overhead across a variety of input shapes and precision settings.

\subsection{FT K-means under Error Injections}
\begin{figure}[ht]
    \vspace{0mm}
    \centering
    \includegraphics[width=\linewidth]{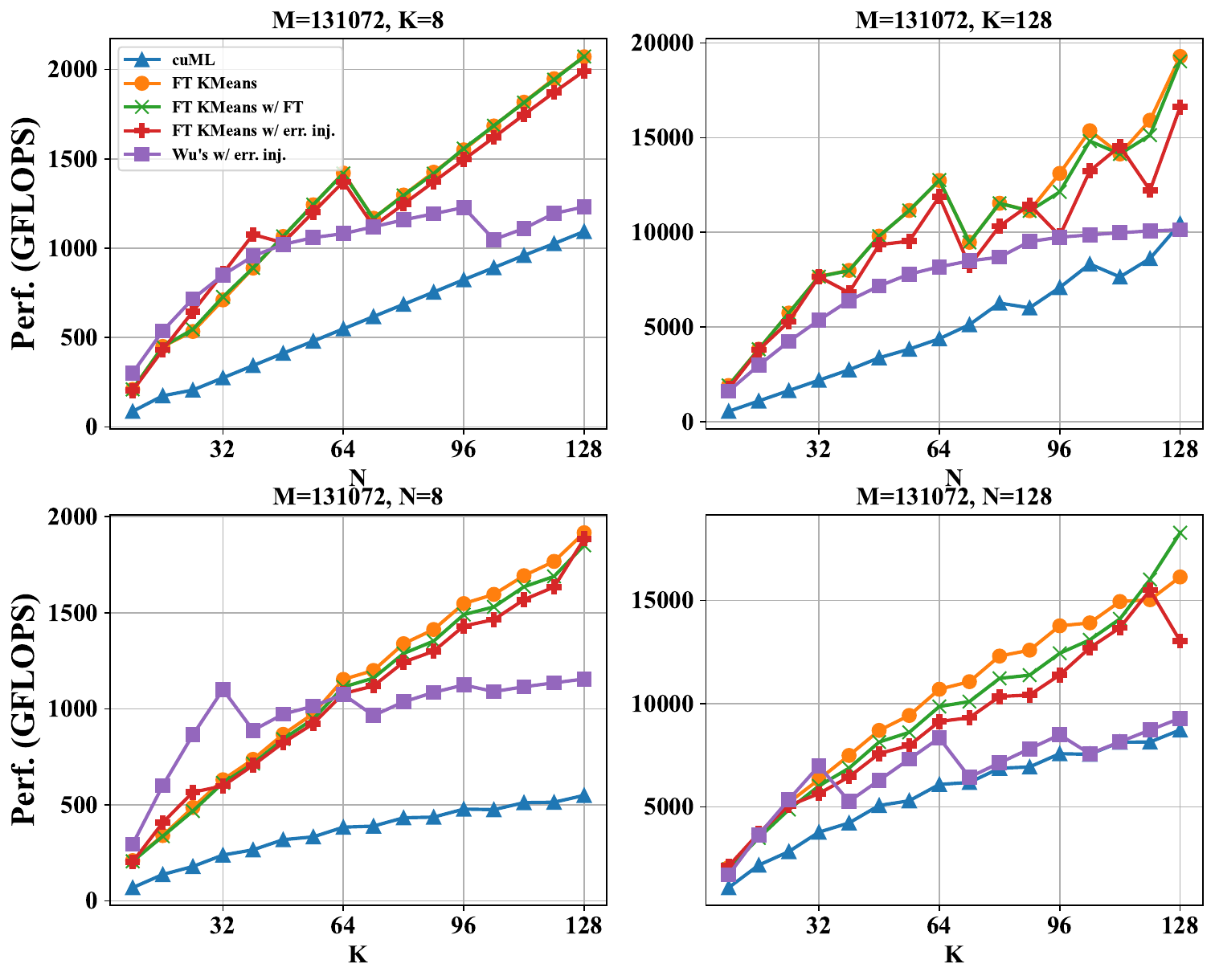}
    \caption{Benchmark FT K-means with error injection: FP32.}
    \label{fig:a100_fixMK_fp32_err}
    \vspace{0mm}
\end{figure}

As illustrated in Figure \ref{fig:a100_fixMK_fp32_err}, the experimental evaluation of the FT K-means algorithm under error injection reveals a minimal average overhead of approximately 2.36\%, demonstrating the algorithm's efficiency in handling errors with minimal additional computational cost. The results indicate robust performance across various input shapes, with overhead percentages ranging from a slight reduction of about -0.93\% to a modest increase up to 9.49\%. This low overhead highlights the effectiveness of the fault tolerance mechanisms integrated into the FT-KMeans, ensuring correctness even in the presence of injected errors. Wu's FT scheme introduces an overhead of 30\% due to a suboptimal GEMM baseline without using the asynchronous memory copy.

\begin{figure}[ht]
    \vspace{0mm}
    \centering
    \includegraphics[width=\linewidth]{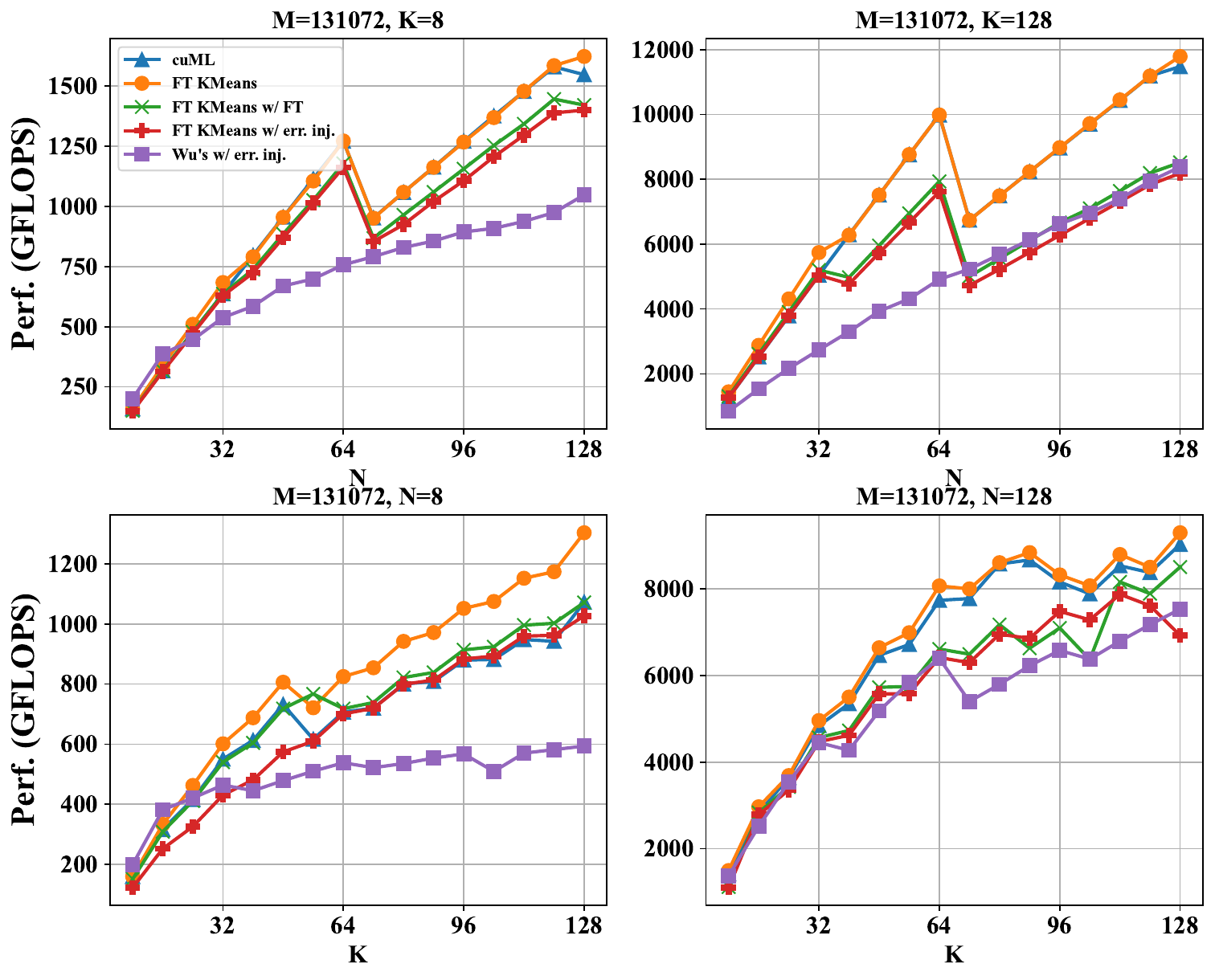}
    \caption{Benchmark FT K-means with error injection: FP64.}
    \label{fig:a100_fixMK_fp64_err}
    \vspace{0mm}
\end{figure}

 In Figure \ref{fig:a100_fixMK_fp64_err}, the evaluation of the FT-KMeans algorithm with error injection in a 64-bit floating point (fp64) environment shows a low average overhead of approximately 9.21\%s. The data demonstrates that configurations with fewer features (N=8 and N=128) exhibit low overheads of 0.79\% and 0.84\% respectively, highlighting the algorithm's effective fault tolerance mechanisms which ensure minimal performance degradation even in high-precision settings. Meanwhile, fixed K scenarios, K=8 and K=128, show higher overheads of 10.12\% and 24.07\%, indicating increased computational complexity under fault conditions. 

\subsection{Performance Evaluation on T4}

\begin{figure}[ht]
    \vspace{0mm}
    \centering
    \includegraphics[width=\linewidth]{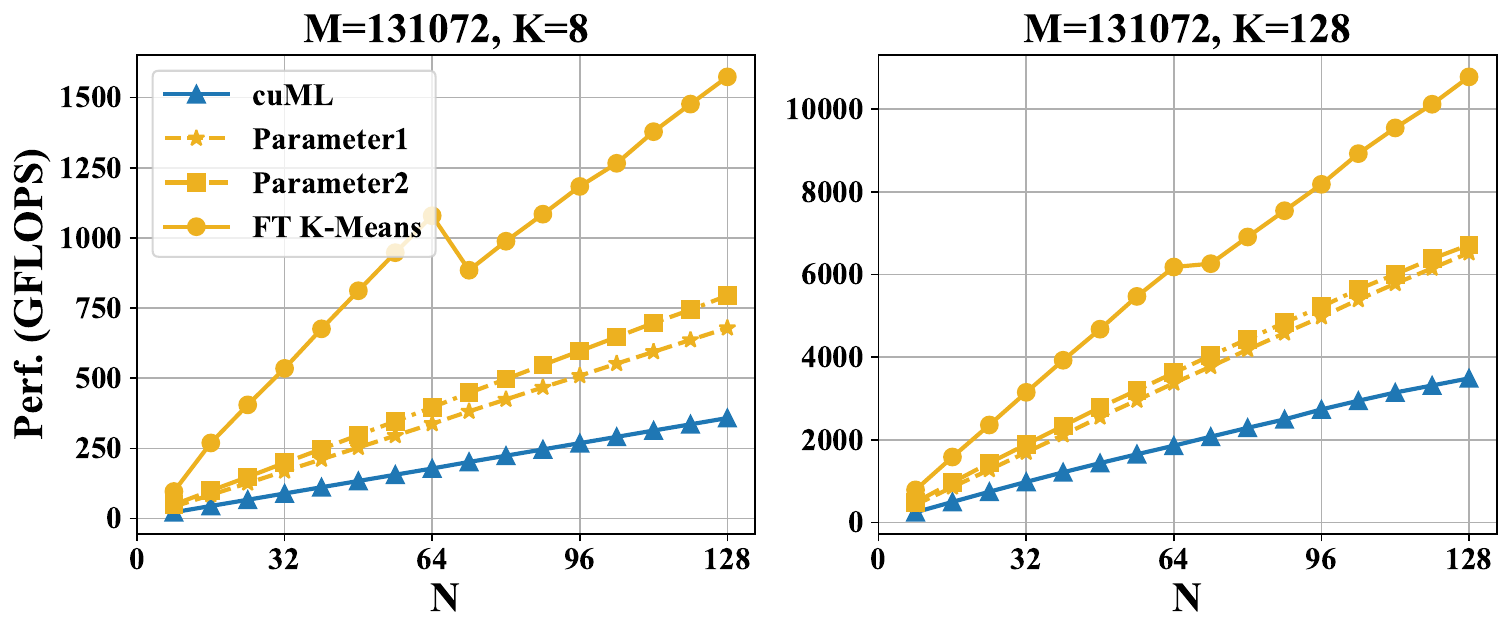}
    \caption{FP32 precision comparison of K-means performance at distance step without fault tolerance with FT K-means, Selected parameters, and cuML on a T4 GPU, with M and K fixed.}
    \label{fig:T4_fixMK_fp32}
    \vspace{0mm}
\end{figure}

Figure \ref{fig:T4_fixMK_fp32} demonstrates a performance evaluation in distance step between FT K-means, two selected parameters relative to cuML (all without fault tolerance) in FP32 precision when M and K are fixed. The definitions of Parameter1 and Parameter2 are similar to our evaluation of A100. And they are consistent with the values on A100 to the greatest extent. However, they have better performance compared to cuML's parameter in this architecture, with speeds up to 184\% and 208\%. Using code generation strategy, we achieved 413\% speedup compared to cuML under FP32 precision, and the gain in performance is significant even when K is relatively larger.

\begin{figure}[ht]
    \vspace{0mm}
    \centering
    \includegraphics[width=\linewidth]{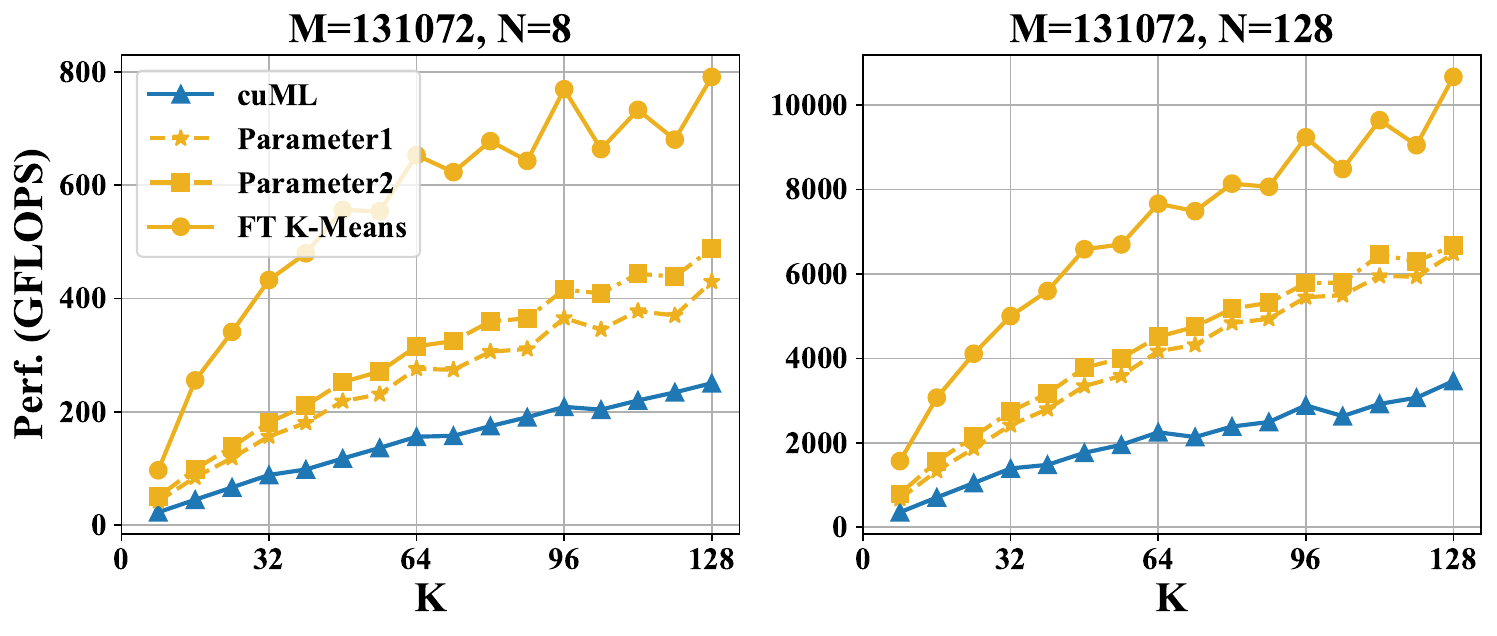}
    \caption{FP32 precision comparison of K-means performance at distance step without fault tolerance with FT K-means, Selected parameters, and cuML on a T4 GPU, with M and N fixed}
    \label{fig:T4_fixMN_fp32}
    \vspace{0mm}
\end{figure}

Figure \ref{fig:T4_fixMN_fp32} offers a performance evaluation in distance step between FT K-means, two selected parameters relative to cuML (all without fault tolerance) in FP32 precision when M and N are fixed. The main parameter settings are similar to the previous section, and $N=8$ and $N=128$ indicate fewer clustering centroids and relatively more clustering centroids. The selected parameters achieve 183\% and 206\% speed up against cuML. Using code generation strategy, we obtained 381\% speedup compared to cuML under FP32 precision, which is similar to fixing M and K.

\begin{figure}[ht]
    \vspace{0mm}
    \centering
    \includegraphics[width=\linewidth]{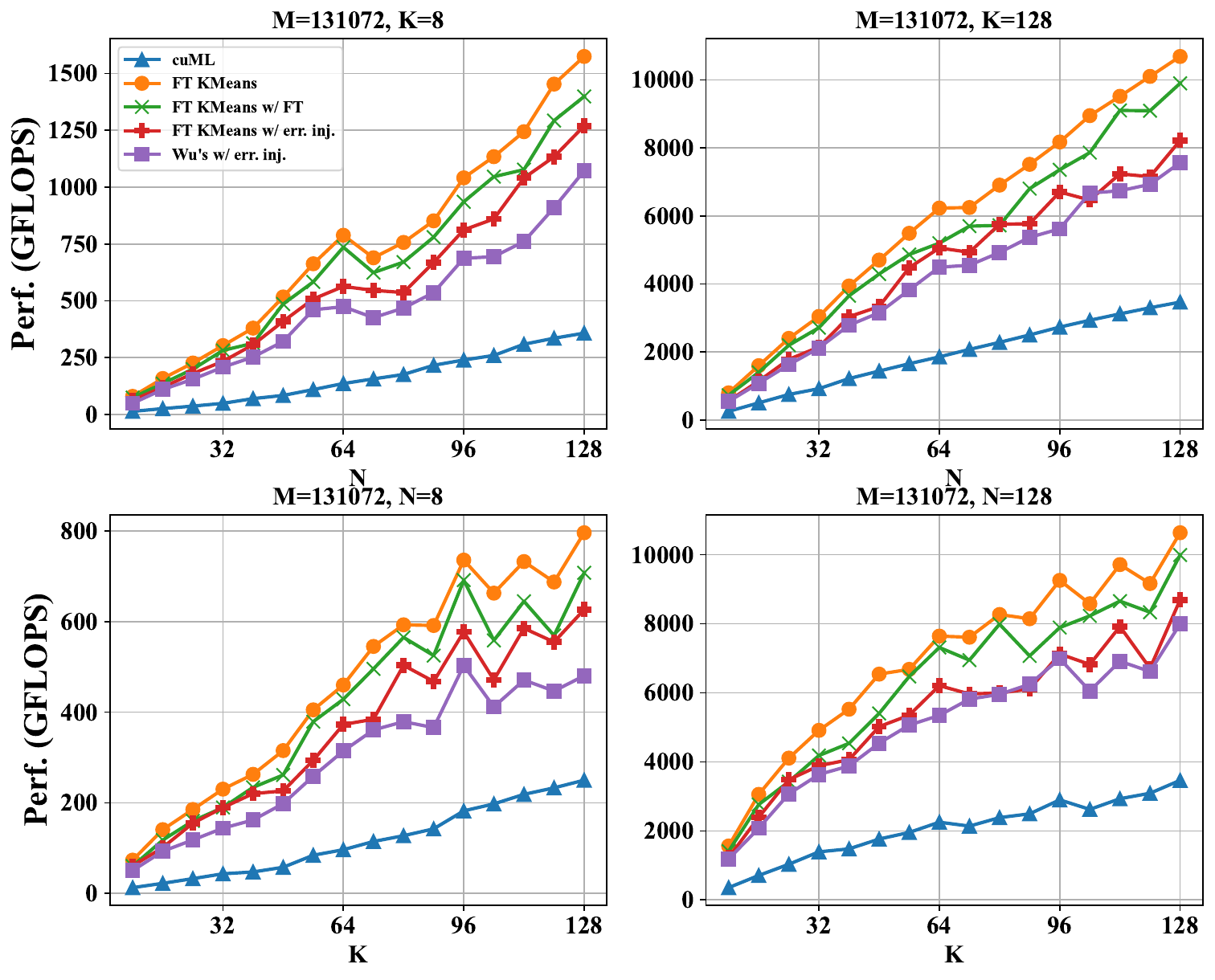}
    \caption{Benchmark FT K-means with error injection on T4: FP32.}
    \label{fig:t4_fixMK_fp32_err}
    \vspace{0mm}
\end{figure}

Figure \ref{fig:t4_fixMK_fp32_err} benchmarks the performance of FT K-means with or without fault tolerance for single precision. FT K-means shows an average overhead of $18\%$ with fault tolerance and $30\%$ under error injection. Compared with Wu's ABFT, FT K-means has a $60\%$ improvement due to the elimination of threadblock-level synchronization.

%% file: Sections/Section6-Conclusion.tex
\section{Conclusion}

 In this paper, we introduce FT K-means, a high-performance GPU-accelerated implementation of K-means with online fault tolerance. We first present a stepwise optimization strategy that achieves competitive performance compared to NVIDIA's cuML library. We further improve FT K-means with a template-based code generation framework that supports different data types and adapts to different input shapes. A novel warp-level tensor-core error correction scheme is proposed to address the failure of existing fault tolerance methods due to memory asynchronization during copy operations. Our experimental evaluations on NVIDIA T4 and A100 GPUs demonstrate that FT K-means without fault tolerance outperforms cuML's K-means implementation, showing a performance increase of 10\%-300\% in scenarios involving irregular data shapes. Moreover, the fault tolerance feature of FT K-means introduces only an overhead of 11\%, maintaining robust performance even with tens of errors injected per second.

\section*{Acknowledgement}

This work was supported by the U.S. Department of Energy, Office of Science, Office of Advanced Scientific Computing Research, and Scientific Discovery through the Advanced Computing (SciDAC) program under Award Number DE-SC0022209.